\documentclass[12pt,preprint]{aastex}

\newcommand{\kms}{km~s$^{-1}$}
\newcommand{\halp}{H$\alpha$}
\newcommand{\msun}{\rm M_{\odot}}
\newcommand{\etal}{et~al.~}

\begin{document}
\title{Kinematic Structure of the Orion Nebula Cluster
and its Surroundings\altaffilmark{1}}

\author{G\'{a}bor F\H{u}r\'{e}sz\altaffilmark{2,3,4},
Lee W. Hartmann\altaffilmark{5},
S. Thomas Megeath\altaffilmark{6},
Andrew H. Szentgyorgyi\altaffilmark{2},
Erika T. Hamden\altaffilmark{7}
}

\altaffiltext{1}{Observations reported here were obtained at the MMT Observatory, a 
joint facility of the Smithsonian Institution and the University of Arizona}
\altaffiltext{2}{Center for Astrophysics, 60 Garden Street, Cambridge, MA 02138}
\altaffiltext{3}{most of this work was done while GF was a research fellow of the Konkoly Observatory of the Hungarian Academy of Sciences,
P.O. Box 67, H--1525 Budapest, Hungary}
\altaffiltext{4}{University of Szeged, Department of Experimental Physics, 
Dom ter 9, H--6723 Szeged, Hungary}
\altaffiltext{5}{Dept. Astronomy, University of Michigan, 500 Church St., 830 Dennison Building, Ann Arbor, MI 48109}
\altaffiltext{6}{Dept. Astronomy, University of Toledo, 2801 West Bancroft Street, Toledo, OH 43606}
\altaffiltext{7}{Dept. Astronomy, Harvard University, Cambridge, MA}

\begin{abstract}
We present results from 1351 high resolution spectra of 1215 stars in the 
Orion Nebula Cluster (ONC) and the surrounding Orion 1c association,
obtained with the Hectochelle multiobject echelle spectrograph on the 6.5m 
MMT. We confirmed 1111 stars as members, based on their radial velocity and/or
\halp~ emission. The radial velocity distribution of members
shows a dispersion of $\sigma=3.1 $ \kms.
We found a substantial north-south velocity gradient and spatially coherent structure in
the radial velocity distribution, similar to that seen in the molecular gas in the region. 
We also identified several binary and high velocity stars, a region exhibiting signs of 
triggered star formation, and a possible foreground population of stars somewhat
older than the ONC.  Stars without infrared excesses (as detected with the IRAC instrument
on the {\em Spitzer Space Telescope}) exhibit a wider spread in radial velocity than
the infrared excess stars; this spread is mostly due to a blue-shifted population of
stars that may constitute a foreground population.  We also identify some accreting
stars, based on H$\alpha$, that do not have detectable infrared excesses with IRAC,
and thus are potential transitional disk systems (objects with inner disk holes).
We propose that the substructure seen both in stellar and gaseous component is
the result of non-uniform gravitational collapse to a filamentary distribution of gas.
The spatial and kinematic correlation between the stellar and gaseous components suggests
the region is very young, probably only $\sim1$ crossing time old or less to avoid
shock dissipation and gravitational interactions which would tend to destroy the
correlation between stars and gas.
\end{abstract}


\section{Introduction}

Because most stars are formed in clusters (e.g., \citet{lada03}), 
the processes responsible for cluster formation are important to 
include in any consideration of the mechanisms of star formation.
Observations of very young clusters can provide clues to the initial
conditions of cluster formation if the cluster has not dynamically relaxed.
The increasing sensitivity of infrared studies with both ground-based 
instruments and the Spitzer Space Telescope has made it more feasible
to search for substructure in embedded populations, as seen for example 
in the young cluster NGC 2264 \citep{teix06}.  In addition, gas and dust can
be cleared away rapidly by stellar energy input, revealing many young cluster
members at optical wavelengths.  In the case of NGC 2264, spatially-coherent
kinematic substructure in the stellar population now has been detected using
optical spectroscopy by \citet{fure06}.

The Orion Nebula Cluster (ONC) is a touchstone for studies of cluster formation,
as it is the closest, relatively populous ($\gtrsim 2000$ members) young cluster
containing an O star.  While many stars in the ONC are heavily extincted
(\citet{ali95}; \citet{mcca94}; Carpenter, \citet{carp01}; \citet{mcca02};
see also \citet{odel01} and references therein), evaporation of molecular gas by the
O6-7 central star $\theta^1$~C Ori has revealed many of 
the members of the ONC at visible wavelengths.  
Optical studies of the stellar population show that it is quite young, with
a median age of roughly 1 Myr or less \citep{hill97}, suggesting that it is reasonable
to search for substructure in the cluster related to the initial conditions of formation.

Hillenbrand \& Hartmann (1998, HH98 hereafter) conducted a 
preliminary study of the structure and kinematics of the ONC.
HH98 constructed azimuthally-averaged stellar source counts and showed that
they could be fit with a simple, spherically-symmetric, single-mass King cluster
model of core radius $\sim 0.2$~pc and a central density of 
$\sim 2 \times 10^4$~stars~pc$^{-3}$ (see also \citet{mcca02}).
However, HH98 pointed out that the cluster is not circular in projection; rather, it is
elongated north-south along the same direction as the molecular gas filament,
suggesting that the cluster might not be completely relaxed.

This elongation is particularly apparent on scales larger than 0.5 pc from the center.
The filamentary structure is well expressed in CO observations of the region \citep{ball87},
as well as in the distribution of pre-main sequence (PMS) stars, shown by recent
infrared (IR) observations (Megeath et al 2007, in preparation).
According to the models of \citet{burk04}, self-gravitationally collapsing
gaseous sheets can form such elongated, filament-like clouds.
The numerical simulations show that density and velocity
dispersion becomes larger at the ends, inducing star formation to start
at the tip of the filament, just like the position of ONC within
the Orion A cloud. The collapsing sheet model
also predicts an overall velocity gradient along the filament, in agreement
with CO observations of ONC \citep{ball87}. For such a young star forming
region as ONC, the stars are still very close to their birthplace; therefore,
the stellar kinematics may reflect initial conditions rather than being relaxed.

We therefore conducted a radial velocity (RV) survey of more than 1200 stars
in the northern region of the Orion A cloud to search for clues to the initial conditions
of the ONC's formation.  In this paper we present results based on
high-resolution spectroscopic observations in the $\lambda \sim 6560$~\AA~ region,
which allowed us to measure RVs and identify T Tauri stars (TTS) based on their \halp~ emission.
We find that the stellar population exhibits a strong spatial and kinematic correlation
with the $^{13}$CO gas in the region; the observed substructure conclusively demonstrates
that the ONC is not relaxed, but instead reflects the initial conditions involved in
its formation.  We also comment on the
detailed structure, distributional differences seen among stellar groups
distinguished by infrared, \halp, and spatial and kinematic properties.
In a subsequent paper we will incorporate followup spectroscopy we have obtained
in the Mg triplet region (centered at 5225 \AA) to examine rotational velocities and
spectral typing, along with observations of the 6708 \AA~ Li line to use as an age estimator.

\section{Target Selection}

To conduct the kinematic study we selected targets in the northern end of Orion A. 
Thanks to the dense molecular cloud there are few background
stars in any stellar catalog of the region.
At the same time, the ONC region exhibits a high density or pre-main sequence objects
on the sky so the foreground contamination is relatively low.
Therefore we initially used a simple color-magnitude selection from the 2MASS
catalog to draw the first target list for our observations.

Our intent was to observe multiple fields to cover at least a $\sim1\times2$ degree
area. Previous experience with Hectochelle suggested we could reach a S/N ratio
of 10 or higher for most of our targets in 1 hour exposure time if
stars with $11.5<J<13.5$.
We also initialliy selected stars with $0.2<(H-K)<0.5$ to avoid heavily-reddened
objects, but this also eliminated many stars with disks (see following paragraph).
These selection criteria above therefore resulted in 2319 targets
within a $1.5\times2.7$ degree field centered at
$\alpha = 5^{h}35^{m}$ and $\delta = -5^{\circ}44'$.
This region includes the Trapezium and areas to the south
(NGC 1980) and to the north (NGC 1973--1975--1977, NGC 1981) of the ONC.

\begin{figure}
\includegraphics[angle=270,scale=0.6]{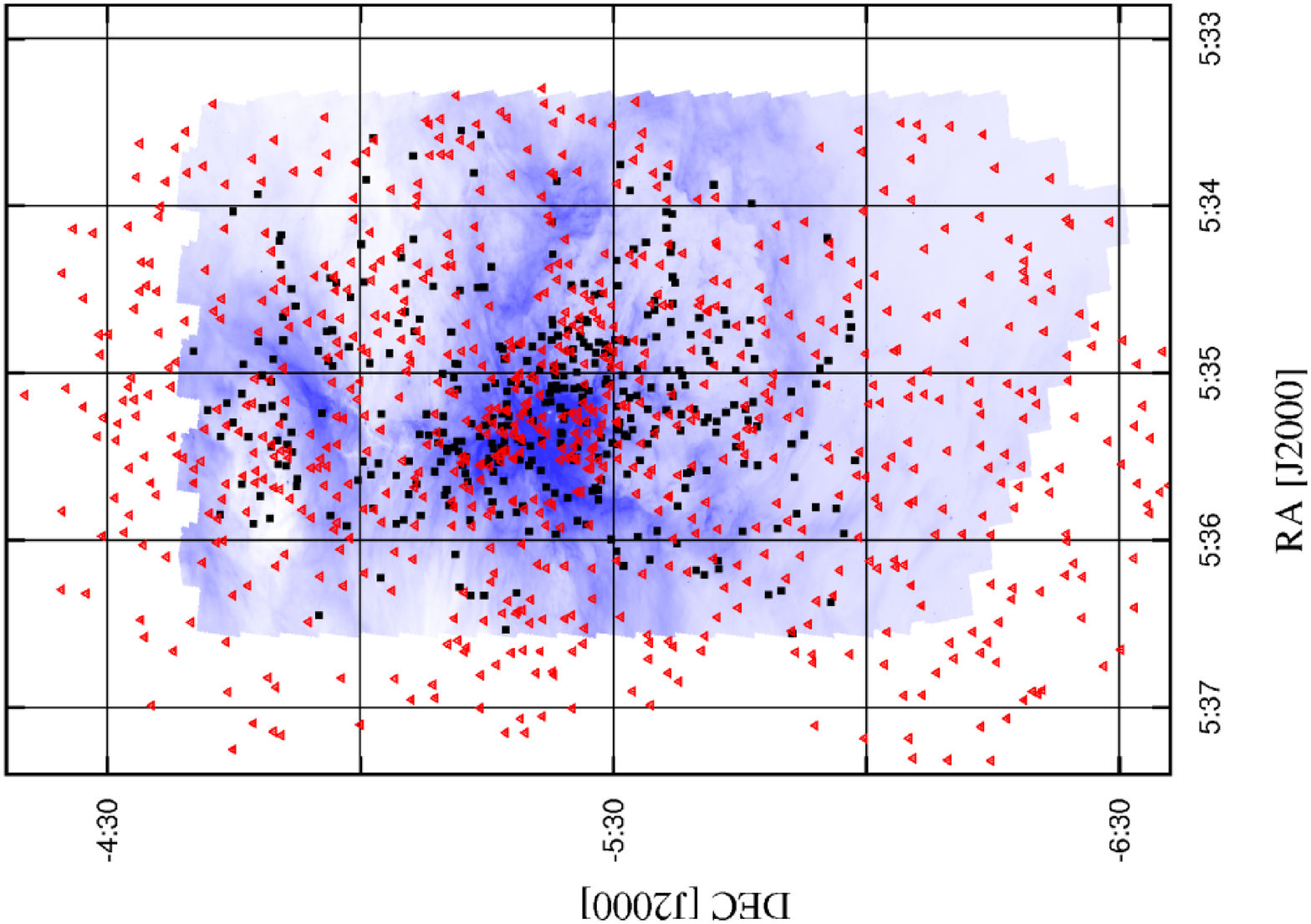}
\caption{Positions of our targets overplotted a IRAC 8$\mu$m mosaic image of Orion Nebula. The open
triangles are stars selected using a 2MASS color-magnitude diagram of the region, the filled
squares are targets from Spitzer/IRAC photometry. See text for details. (The triangle symbols
were brightened over the central region of ONC, in order to increase contrast against the
gray-scaled gas.)
\label{fig:targets}}
\end{figure}

We added young stellar objects to the sample
based on their IR excess as identified using the IRAC and MIPS instruments on board
the {\em Spitzer Space Telescope} (Megeath et al. 2007).
This selection went down
to a fainter magnitude limit of $J=15$, and included additional Class II 
members of the star forming region. These targets were observed during
the second of the two spectroscopic runs, while 2MASS selected
objects were observed in the first. See Figure \ref{fig:targets}
for the spatial distribution of selected targets.
Details will be given in a forthcoming paper (Megeath et al. 2007).

\section{Spectroscopic Observations}
\label{observations}

The spectroscopic observations were carried out using the Hectochelle 
multi-object spectrograph \citep{sain98} 
at the 6.5m MMT telescope in Arizona. 
We used the 190 \AA~ wide echelle order centered at \halp, because we could get RV information
as well as record the stellar hydrogen emission profiles, a sign of youth
and membership, at the same time.  This is
the same instrumental setup that was
used for the observations reported in \citet{fure06} (see section 2.2 of that
paper for details; also \citet{auro05}). 
 
The first set of spectra was taken in 2004 December, when a total of 866 stars
in 4 fields were observed (see Table \ref{table:obs}). Initially we chose not to bin
in the resolution direction, which resulted in 
low signal--to--noise (S/N) values for the faintest targets. 
Therefore, we used a $2\times2$ binning in 2005 November in order to decrease the readout noise
in the data and thus go fainter.
Because Hectochelle oversamples the resolution element (the PSF of a
resolved spectral feature has $\sim3.5$ pixel FWHM), this binning does not result in a serious
loss or resolution.  Accordingly, the accuracy of RVs
derived for the 485 stars observed in 2005 are comparable to the 2004 values.

During the two observing run we collected a total of 1351 spectra of
1215 stars within 7 Hectochelle fields. The location of these objects
are shown in Fig \ref{fig:targets}, overplotted a color IRAC mosaic image
covering most of the $1\times2.5$ degree region explored by spectroscopy.

Sky subtraction would have been very difficult in the vicinity of ONC, as
the nebular emission is so strong and can vary rapidly on small
spatial scales. Consequently, placing a few sky sampling fibers over a
field is not adequate. The best approach would be to take an
offset sky exposure, where all the fibers remain at the target locations
and the telescope is moved by a few arcseconds to set targets off the
fibers, resulting in sampling the background right next to each targets.
However, this procedure would double the exposure
time; due to time constraints we did not take such exposures. 
We simply excluded the emission features not to be important for RV measurements, 
which did not result in an appreciable loss of absorption features 
used in the cross-correlations.

\section{Data Reduction}
\label{datared}

The spectra were extracted, calibrated and normalized by an automated, IRAF
\footnotetext[1]{IRAF (Image Reduction and
Analysis Facility) is distributed by the National Optical Astronomy
Observatories, which are operated by the Association
of Universities for Research in Astronomy, Inc., under contract with
the National Science Foundation.} based pipeline, utilizing the
standard spectral reduction packages and tools.  Some of the
\halp~ emission was so strong that a given aperture became partially
saturated, and this affected the \halp~ profile of neighboring
apertures. As the RV determination is based on absorption lines
outside of the \halp~ region, it usually did not cause problem
in measuring RV as we excluded any emission portion of the spectrum
from the cross correlation.
Sky subtraction was not performed due to the lack of
proper background sampling, as mentioned above, so the stellar
\halp~ and profiles contain nebular emission (N[II]) is basically all nebular).

The velocities were derived by cross-correlating (CC) each observed
spectrum with a set of templates, using the \textit{rvsao}
package within the IRAF environment. As we found in \citet{fure06} by
finding the most similar template to a given object in such
``multi template method'' yields a more accurate RV. Instead of using
observed templates, for this present work we adopted synthetic
spectra from the library of \citet{muna05}.
Even though the unbinned
observations of 2004 yield higher resolution ($R\sim38,000$)
than the $R=30,000$ for the synthetic ones, downgrading the
resolution of the data for the CC was found unnecessary. 
Comparisons between the cross-correlation functions (CCF)
of the original and smoothed (resolution downgraded to
match $R\approx30,000$) data against the same templates
showed no differences in the measured RV value larger than
the estimated error in fitting the peak of the CCF.
For the spectra recorded in 2005 there was an even better match
between the resolutions of the data and the templates,
due to the $2\times2$ binning.

The set of templates was a 3 dimensional grid in effective
temperature ($3500<T_{eff}<6000$, with $250 K$ steps),
surface gravity ($1.0<log(g)<5.0$, with steps of $0.5$)
and rotational velocity (0,2,5,10,15,20,30,40,50,75 an 100 \kms).
The metallicity was chosen to be $[M/H]=+0.5$, as
a possible one for young stars in the solar neighborhood,
and its value was fixed because of the common origin of our targets.
(Other variables of the full original library were set to:
micro-turbulence $K=2$ \kms; no [$\alpha$/Fe] enhancement
and using the new ODF models). A total of 1089 templates
were compared to each spectrum, and the following values
were determined:
\textit{S}, the height of the CCF;~\textit{R}, the signal-to-noise
of the CC (for details see \citet{fure06}, \cite{kurt98} and
\citet{tonr79});~ \textit{$V_{rad}$}, the heliocentric-corrected radial
velocity;~\textit{$V_{err}$}, the error of radial velocity determination.
These values are given in Table \ref{table:hecto_targets_mem} and
\ref{table:hecto_targets_nonmem} along with other result of
spectroscopy and photometry.

For each observed spectrum the CCF values of \textit{S}, \textit{R} and
$V_{rad}$ were plotted
and the global peak was localized for \textit{S}. The parameters of this
best fitting template were adopted as the astrophysical parameters
of the given star, but only in case if \textit{S}, \textit{R} and
$V_{rad}$ were changing smoothly within a small range
with varying template parameters (as expected for a robust result).
We found the variation of these CCF parameters (S, R and $V_{rad}$)
to be a good indicator of noise, as low S/N ratio and/or featureless
spectrum yield very different CCFs for different templates, and exhibit no global 
(outstanding) CCF peak but
several local peaks of the same height (0.1-0.2). 
Significant CCFs yield global peaks at least 30-50\% higher than any other 
noise peak. 

We plotted the CCF for the best matching template for each star, and
examined the result by eye to make sure the global peak identified did not exhibit
any sign of binarity (side-lobes or a secondary peak with comprable
peak height). If a companion was found (or a high probability of a possible
companion was suggested by the CCF), we marked the star as a binary (possible
binary; see Table \ref{table:hecto_targets_mem} and \ref{table:hecto_targets_nonmem}).

As a result there were 1049 stars 
out of 1215 yielding an $R>4$ (only these were used in the later analysis),
with a mean R value of 13.7. As a comparison, using 10 actually
observed templates (the ones from \citet{fure06}) we only obtained 547
stars with $R>4$ and a mean R value of only 7.8.  Thus, increasing
the number of templates led to a significant
increase in the number of objects with accurate RV values. 

We found a 0.8 \kms~ offset between the mean heliocentric velocities of the 2004 and 2005 observations.
This is likely due to the different (improved) calibration system used in 2005.
The shift was determined by comparing velocities of 35 stars observed in both runs
with R values larger than 8. The values are listed in Table \ref{table:hecto_targets_mem}
and \ref{table:hecto_targets_nonmem} are corrected for this zero point offset,
by shifting the 2004 values to the 2005 scale.

As discussed by Tonry \& Davis (1979), the RV error 
($\Delta V_{rad}$ in Table \ref{table:hecto_targets_mem}) is expected to vary as
\begin{equation}
\Delta V_{rad} \sim W/(1+R)\,,
\end{equation}
with W being a constant related to the velocity width of the CCF.  We estimated
that $W \sim 10$~\kms~ and found this to be in good agreement with multiple observations
of program objects after correcting for the above zero point offset and eliminating
possible binaries.

\section{Results}

\subsection{Velocity Distribution and Correlations Between Stellar and Gaseous Component}
\label{velo}

\begin{figure}
\includegraphics[angle=270,scale=0.6]{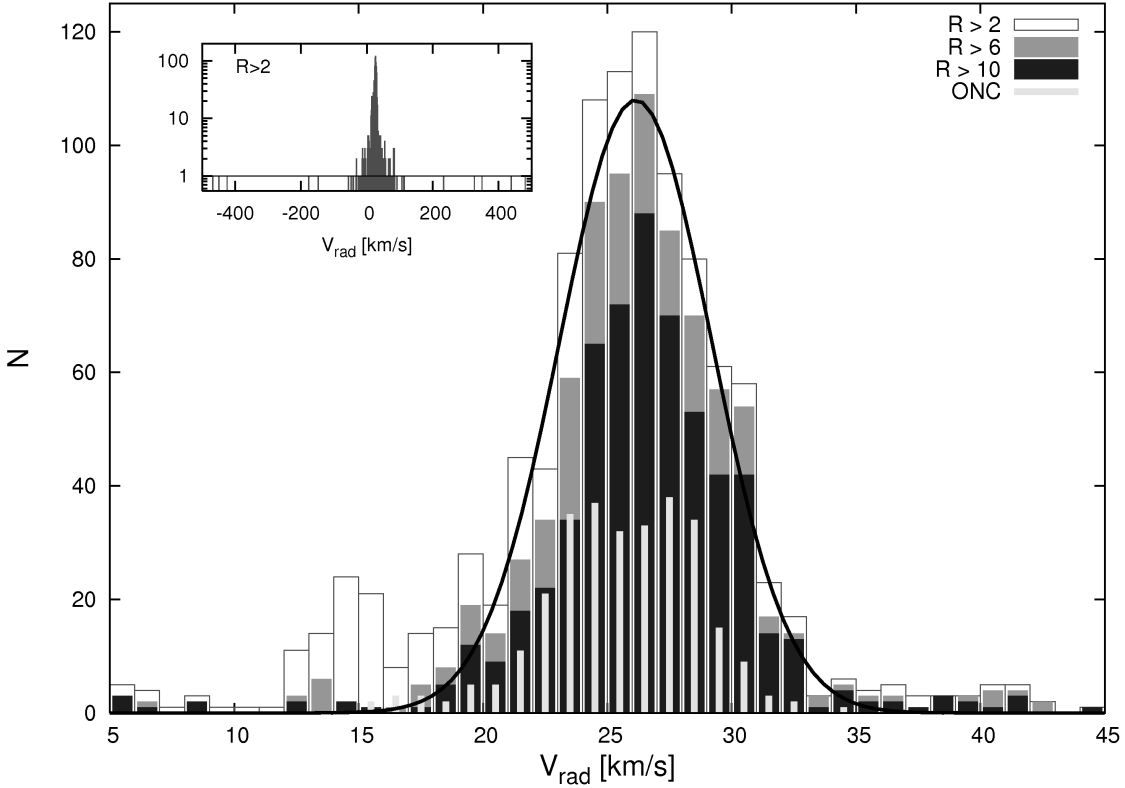}
\caption{Histogram of measured heliocentric radial velocities, displayed for four selections of
stars. If the faintest spectra are included, which also yield low $S/N$ ratio (R value)
in the cross correlation, a false peak appears in the distribution at the heliocentric
correction velocity ($V_{rad}=14$ \kms). Therefore in the later analysis we
only used $R>4$ spectra, which we found not to be influenced by scattered moonlight.
For the main peak the shape is independent of the R value, and can be fitted with
a $\sigma=3.1$ \kms~ Gaussian, however there are is one relevant departure
from the fit at $V_{rad}=19$ \kms. Distribution of
stars within $15\arcmin$ of Trapezium are displayed with narrower, light-gray
shaded impulses, to show how the distribution of ONC stars compare to the entire region.
Note the more flattened peak, but also the agreement in the mean velocity and
dispersion.
The insert shows the entire 1000 \kms~ wide velocity range explored in the cross correlation,
and even including the lowest R value spectra ($R>2$) the global peak around $V_{rad}=26$ \kms~
is very well defined, with only a very few outliers (these are usually faint, noisy spectra).
\label{fig:rv_hist}}
\end{figure}

The distribution of measured heliocentric radial velocities is shown in Fig. \ref{fig:rv_hist}.
The histogram of the entire sample is displayed for three different S/N ratio of
the cross correlation, and we also show the distribution of a spatially selected sub-sample.
The $R>2$ group includes the faintest spectra, on which scattered moonlight causes the
the cross-correlation to fix on the solar spectrum
rather than the stellar lines; this leads to a false peak near the heliocentric
correction value ($\sim14$ \kms). However, most of the faint spectra follow the distribution
drawn by the higher S/N ratio selections. In any event, we did not
use any spectra with R values less than 4 in the following analysis.

The distribution of the entire $R>4$ sample peaks at a heliocentric velocity of 26.1 \kms~
(see Gaussian fit on Fig. \ref{fig:rv_hist}, in good agreement with our
initial study of ONC \citep{auro05}). The dispersion we found, $\sigma=3.1$ \kms~
(with 10 and 40 \kms cutoff values), 
is somewhat higher than the 2.3 \kms~ value of the previous Hectochelle study,
or the 2.5 \kms~ derived from proper motions \citep{jone88},
but the present sample includes a larger region (see Fig. \ref{fig:targets}).
Nevertheless, spatially selecting a sub-sample in the close vicinity of ONC,
within a $15\arcmin$ radius of Trapezium (292 stars with $R>4$), we still found
a mean velocity of 25.6 \kms~ and a dispersion of $\sigma=3.1$ \kms.
These numbers are very close to values calculated using the entire sample, but the
distribution is different for these two spatial selections, as it is obvious in
Fig. \ref{fig:rv_hist}. While all the observed stars can be fitted by a $\sigma=3.1$ \kms~
Gaussian, the velocity distribution of ONC stars does not seem sufficiently peaked at
the mean velocity.

We computed the cross-correlation between $-500$ to $+500$ \kms~ heliocentric
velocities, but
found only a small number of stars farther from the peak than $\pm4{~}\sigma$. 
Between $4-15{~}\sigma$ there are stars with high R values, but outside of this
region there are only a few, usually very faint, outliers. The insert
in the upper left of Fig. \ref{fig:rv_hist} shows how clean the distribution
is over the explored RV range, even including the $R>2$ spectra. The y axis
is log scaled, to better show the wings of the peak. (A horizontal
line is also drawn at $N=1$.) Based on the morphology of the histogram
we define RV membership for stars with velocities in the $26.1 \pm 4\sigma$
range, or between 13.7 and 38.5 \kms. Even if the RV was out of this range,
or was very uncertain due to low R value, we still considered a star as
member if its spectra exhibited non-nebular (i.e., large velocity width)
\halp~ emission.
For the details on members and non-members
see Tables \ref{table:hecto_targets_mem} and \ref{table:hecto_targets_nonmem}.

\begin{figure}
\includegraphics[angle=270,scale=0.7]{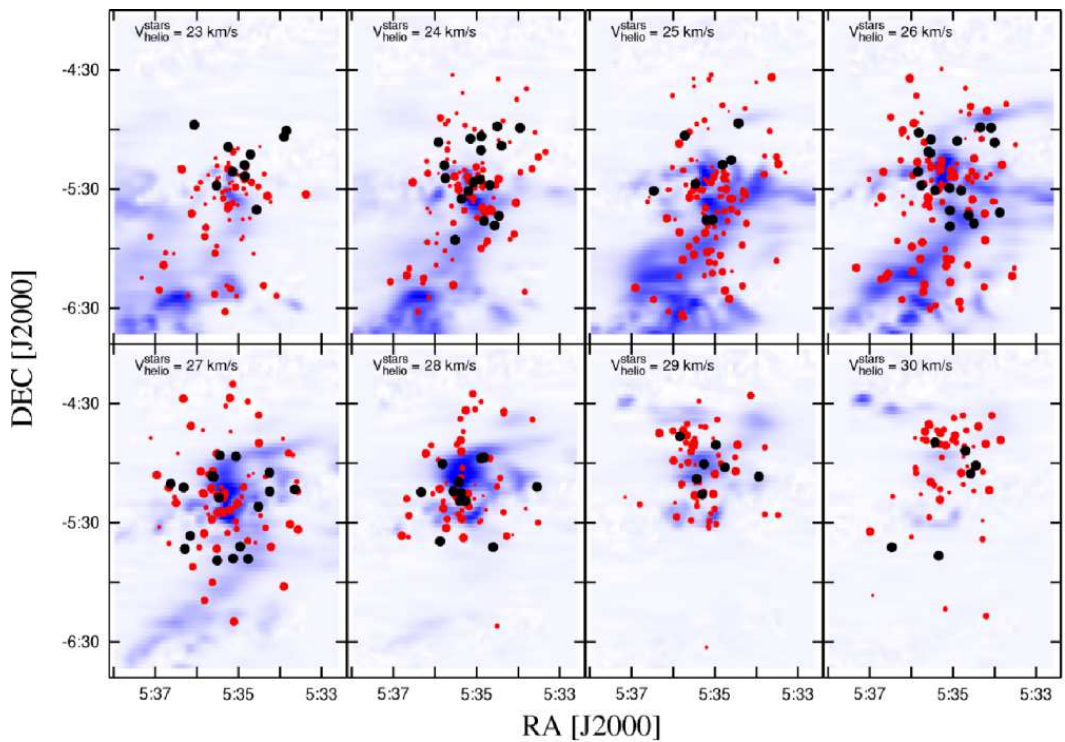}
\caption{Velocity channel maps of the ONC region. The dots represent stars with one (red)
or more (black) RV measurements, while blue is the molecular gas observed in $^{13}CO$,
both displayed in 1 \kms~ channels. LSR velocities of the gas is converted to heliocentric
($V_{helio}=V_{LSR}+17.5$). The size of the dots correspond to the R value, larger dot means
more accurate velocity. The north-south velocity gradient is obvious and the correlation
between the spatial and density distribution of stars and molecular gas is very significant. 
This suggests that the stars are still co-moving with the gas clouds in which they formed.
\label{fig:rv_chanel}}
\end{figure}

\begin{figure}
\includegraphics[angle=270,scale=0.7]{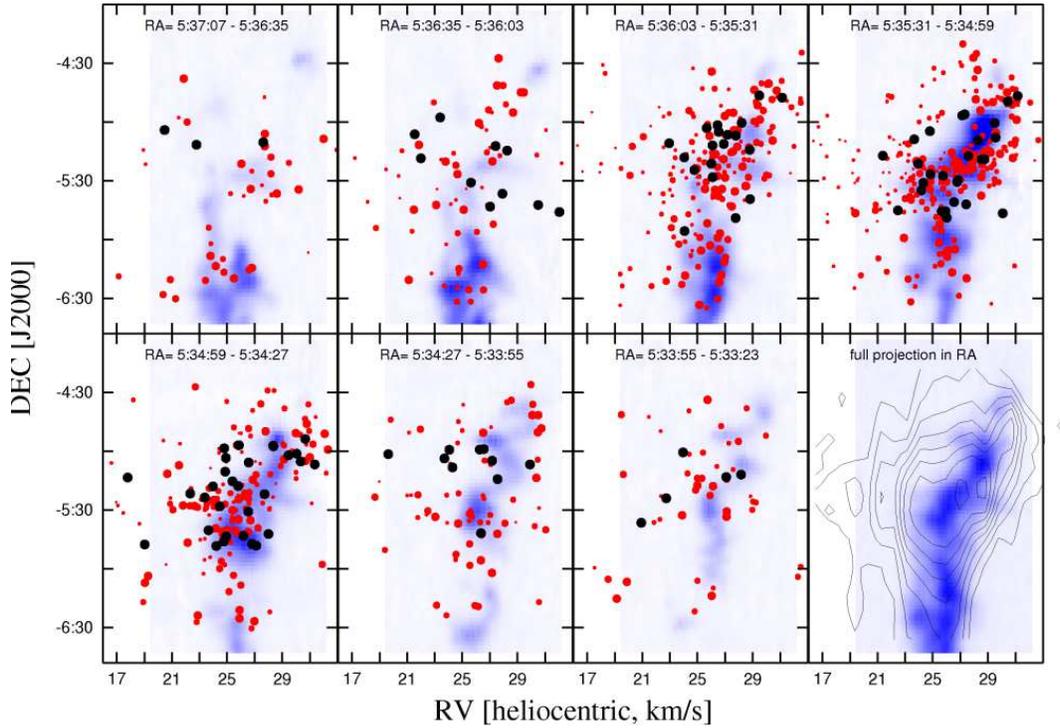}
\caption{Channel maps of the ONC region, in right ascension. Symbols are the same
as for Fig. \ref{fig:rv_chanel}. The approximately 30 second wide RA ranges are noted
in each channel. The correlation between the gaseous and stellar component is most
prominent in the upper right corner, in the channel containing the Trapezium.
On the lower right panel a full projection is shown with stellar density
contours overplotted the gas distribution.
\label{fig:ra_chanel}}
\end{figure}

\begin{figure}
\includegraphics[angle=270,scale=0.7]{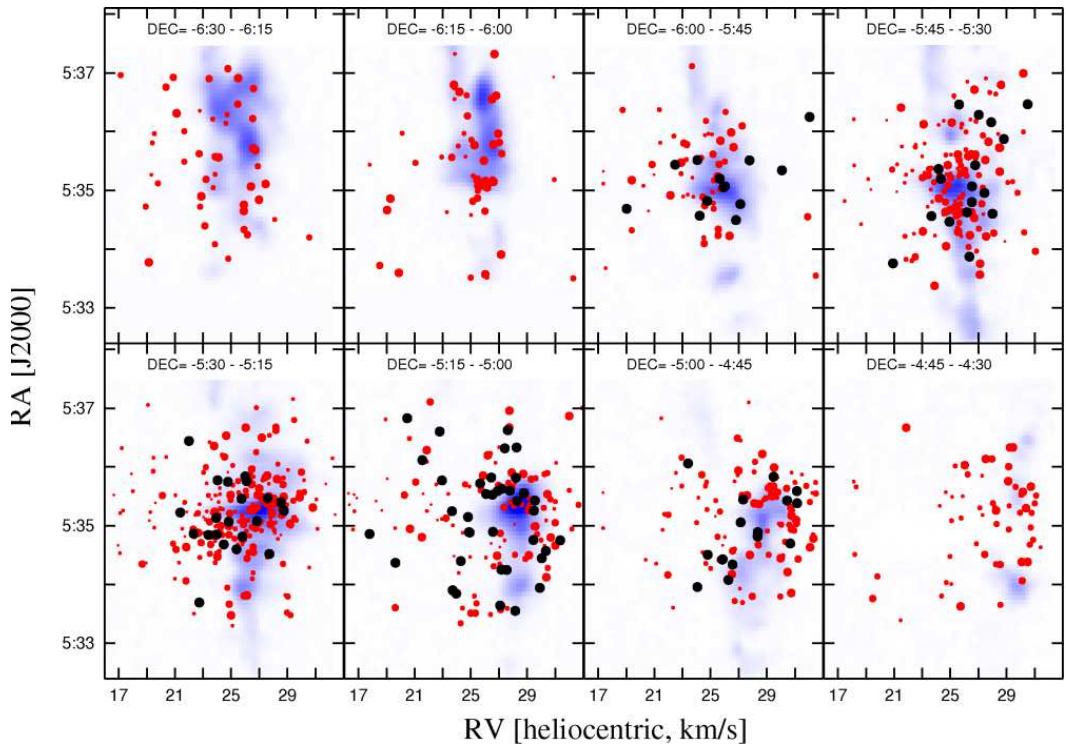}
\caption{Channel maps of the ONC region, in declination. Symbols are the same
as for Fig. \ref{fig:rv_chanel}. The 15 arcminute wide DEC ranges are noted
in each chanel.
\label{fig:dec_chanel}}
\end{figure}

Figure \ref{fig:rv_chanel}, \ref{fig:ra_chanel} and \ref{fig:dec_chanel}
display channel maps, slicing the right ascension--declination--velocity data cube
in RV, RA and DEC, respectively. The stars (filled circles) are shown together with the
molecular gas (blue shading) in order to compare structure between the stellar
and gaseous component. 
The $^{13}$CO measurements of \cite{ball87} were converted
from LSR to heliocentric, to match our RV measurements, by adding 17.5 \kms~ to the LSR
values. The size of stellar symbols are coded with R value, larger circle meaning
higher $S/N$ in the cross correlation and hence more accurate RV measurements in general. As some 
stars were observed more than once, we distinguish these by black color as their
averaged RV values are even more accurate.
All of these maps show an overall north--south
gradient in RV and significant substructure. 

The morphology seen in the stellar
population is very similar to that of the molecular gas. This is most
apparent on Fig. \ref{fig:ra_chanel}, where the projection is in RA, as the
orientation of the molecular gas has a filamentary structure mostly extending along DEC.
To emphasize this structural parallelism, on the last panel of this figure we present
a full projection of the $^{13}CO$ data with contour plots of stellar density.
(For the contours the number of stars was calculated in 1 \kms~ by 0.1$\degr$~ bins
with a grid resolution of 0.5 \kms~ and 0.05$\degr$.)

\subsection{\halp~ Emission profiles}
\label{halpha}

As described earlier (section \ref{observations}), we did not take the observations
that would be necessary for adequate sky subtraction in this region,
where the nebular background can vary strongly over small scales. 
This makes the analysis of \halp~ profiles difficult, but not impossible 
in cases where the stellar
component is much broader than the (relatively narrow) nebular component.
However, nebular emission can be so strong that the stellar signature is
barely visible as asymmetries in the wings, at very low intensity levels.
This means that neither the canonical 10~\AA~ limit in equivalent width
nor the 270 \kms~ limit in full-width at 10\% peak \citep{whit03}
method can be applied on the raw spectra to distinguish between accreting
T Tauri stars (classical T Tauri stars, CTTS) and non-accreting, weak-emission
T Tauri stars (WTTS).

To help identify the broad wings of H$\alpha$, we used a simple
algorithm to identify narrow (FWHM $<10$ \kms) emission peaks
close ($\pm25$ \kms) to the main $^{13}$CO velocity. 
and fit a gaussian component to this peak.
If such local peak was found than it was replaced with a linear segment,
connecting the points where the wider stellar profile started
to deviate from the narrow gaussian component.  While this helped us to
visually inspect the line profiles and make a judgment as to the presence
of broad emission wings, it is obviously not a robust method of eliminating
nebular emission, especially given the complex velocity
structure seen in the densest regions.

\begin{figure}
\includegraphics[angle=270,scale=0.6]{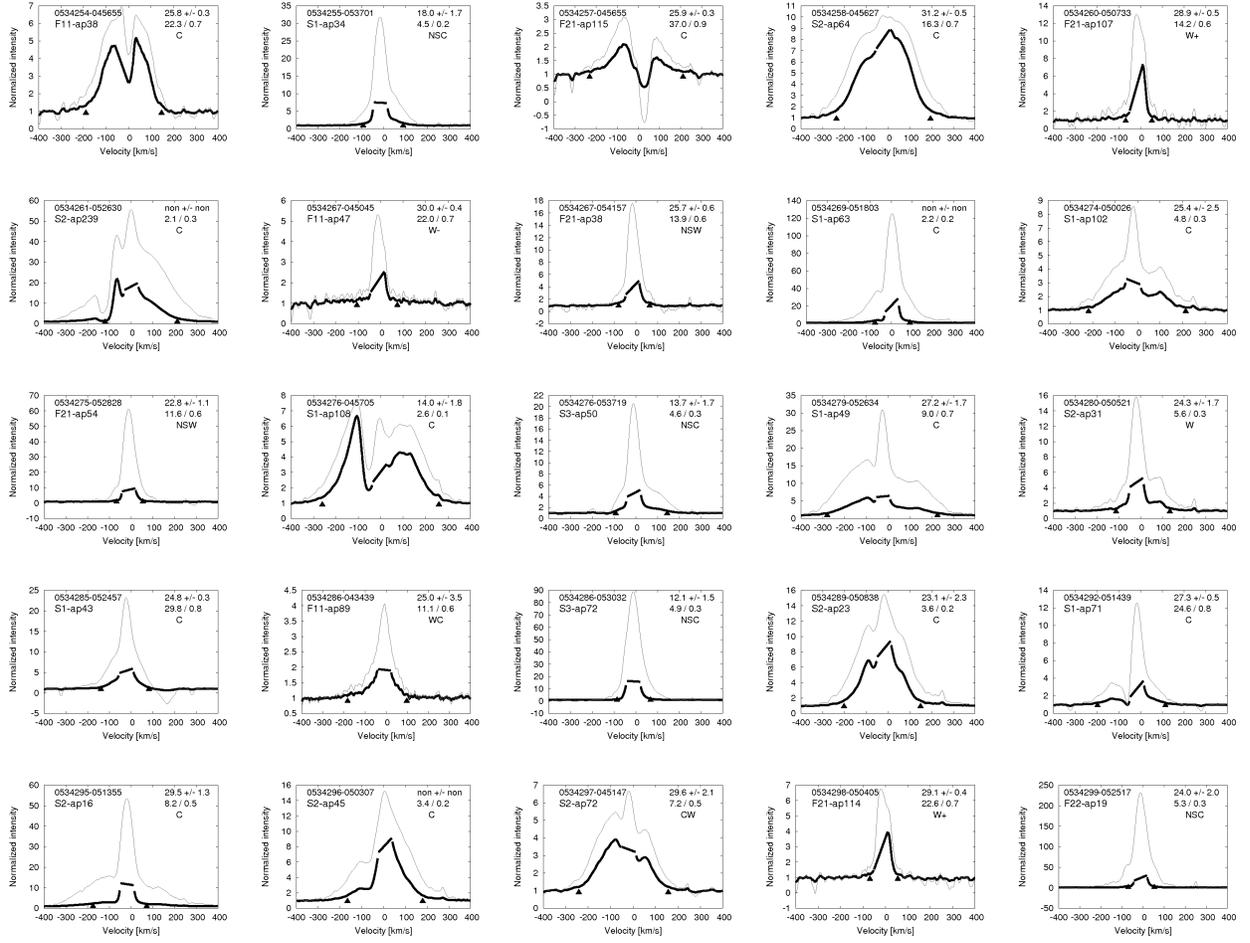}
\caption{Examples of $H\alpha$ profiles for stars with detected emission. Wavelength is converted
to velocity with the zero-point being the RV of the star. (In case of very low R values
the zero-point is set to RV=0 \kms.) 
2MASS and internal identification numbers are given in the upper left corner for each stamp,
while RV and its error, the R and S values, and note on the $H\alpha$ profile are shown 
in the upper right (see Table \ref{table:hecto_targets_mem} and \ref{table:hecto_targets_nonmem}).
The thick solid line is the observed spectrum with linear scaling and with the supposed nebular
component cut off and replaced by a linear segment. The span of the \textit{y} axis 
was scaled to accommodate the peak of the original, unmodified profile, so
one can read the relative strength respect to continuum even the peak was found to 
purely nebular and therefore was cut off. For the original profile, and to
emphasize the sometimes barely visible wings, we also show
as thin solid line a log scaled version of the observed, unmodified profile.
It is scaled to match in peak the original, linear scaled peak height (so the scale
does not apply to this).
Solid triangles mark the 10\% full width for the nebular-line corrected spectrum.
[\textit{See the electronic edition of the Journal for all observed $H\alpha$ profiles
-- or contact the authors}]
\label{fig:halpha_profiles}}
\end{figure}

In cases of strong nebular emission only the low level wings contain
usable information for classification.  We therefore examined the
linear and log scaled plots of each \halp~ profile. A sample of these
plots are presented in Fig. \ref{fig:halpha_profiles}. The linearly scaled,
corrected profile displayed as a bold line, with a linear segment
replacing the supposed and eliminated nebular component. 
The logarithmically scaled version of the uncorrected spectra
is shown as a thin line, and other than emphasizing the wings it also 
makes it the possibile to check if the automated identification and 
subtraction of nebular emission was correct. 
(The intensity values does not apply
to this representation, as it was scaled to fit the linear-scaled 
range described above.) Black triangles mark the full width
at 10\%, which was measured together with EW on the corrected profile.
Those values are listed in in Table \ref{table:hecto_targets_mem}.

We found some contradiction between the two spectroscopy-based classification
scheme of TTSs. For some stars even though the nebular-corrected EW (and sometimes the
uncorrected EW as well) was
significantly smaller than 10 \AA, the full-width at 10\% reached or exceeded 250 \kms,
broadening strongly indicative of accretion (for example: 0535495-042438 = F11\_ap128;
0534257-045655 = F21\_ap115; 0536197-051438 = S1\_ap175; etc.).
With IRAC photometry in hand for several of these stars we were able
to confirm the CTTS status. At the same time in other cases we encountered some
disagreement between very clear CTTS profiles and IRAC photometry
\textit{not} suggesting a disk (see section \ref{discs} below).
In addition, further uncertainties were caused by the artificial subtraction of 
nebular emission component.

Therefore the final notes (classes) on \halp~ profiles
listed in Table \ref{table:hecto_targets_mem} and \ref{table:hecto_targets_nonmem}
are based on a somewhat more complicated classification scheme than CTTS{/}WTTS
or no-emission. In cases of very strong nebular emission with apparently
broad but uncertain wings in H$\alpha$, to note the possible nebular
bias in classification of these stars, we distinguish them with an ``NS''
prefix in the \halp~ note.  A full description of our classification
scheme is given in the notes for Tables \ref{table:hecto_targets_mem}
and \ref{table:hecto_targets_nonmem}.

In terms of statistics, in our particular sample
the total number of CTTS (including C, CD, CW, NSC,
see \halp~ note in Table \ref{table:hecto_targets_mem}) is 581, or 53.6\%,
The total WTTS count is 439 (including W, WC, WD, W+, W-, NSW), or 40.5\%,
for a ratio of $CTTS{/}WTTS=1.35$. 
If only the most confidently classified groups are considered, we have 276 CTTS
(including only C), and 230 WTTS (including 160 W and 70 W+), which 
gives a ratio of $CTTS{/}WTTS=1.20$. 
However, we caution that these numbers {\em cannot} be used to estimate a true
ratio of accreting to non-accreting stars in the region, because our sample
is biased by the selection of infrared-excess stars, as well as being limited in
the sampling of stars in the inner ONC due to crowding.  In addition, our
non-IRAC selection of stars (F-fields) tended to omit stars with large
infrared excess in H-K.

\subsection{Spectroscopic vs. Photometric Disc{/}Accretion Indicators}
\label{discs}

The IRAC color-color diagram $[3.6]-[4.5]$ vs. $[5.8]-[8.0]$  \citep{alle04}
can be used to distinguish between young
stars with and without inner disk emission (and protostars with circumstellar envelopes,
but these were not within the scope of our survey).
In the case of a star having both short-- and long
wavelength infrared excess due to a circumstellar disk, accretion onto the star is
likely taking place, and therefore we expect a broad \halp~ profile,
(the signature of material falling in at high velocity), thus a CTTS
(e.g., White \& Basri 2003).  If the short wavelength excess is 
not present, presumably there is either no disk to accrete or no
inner disk; in most cases, such systems are not accreting and thus
are WTTS.  

\begin{figure}
\includegraphics[angle=270,scale=0.6]{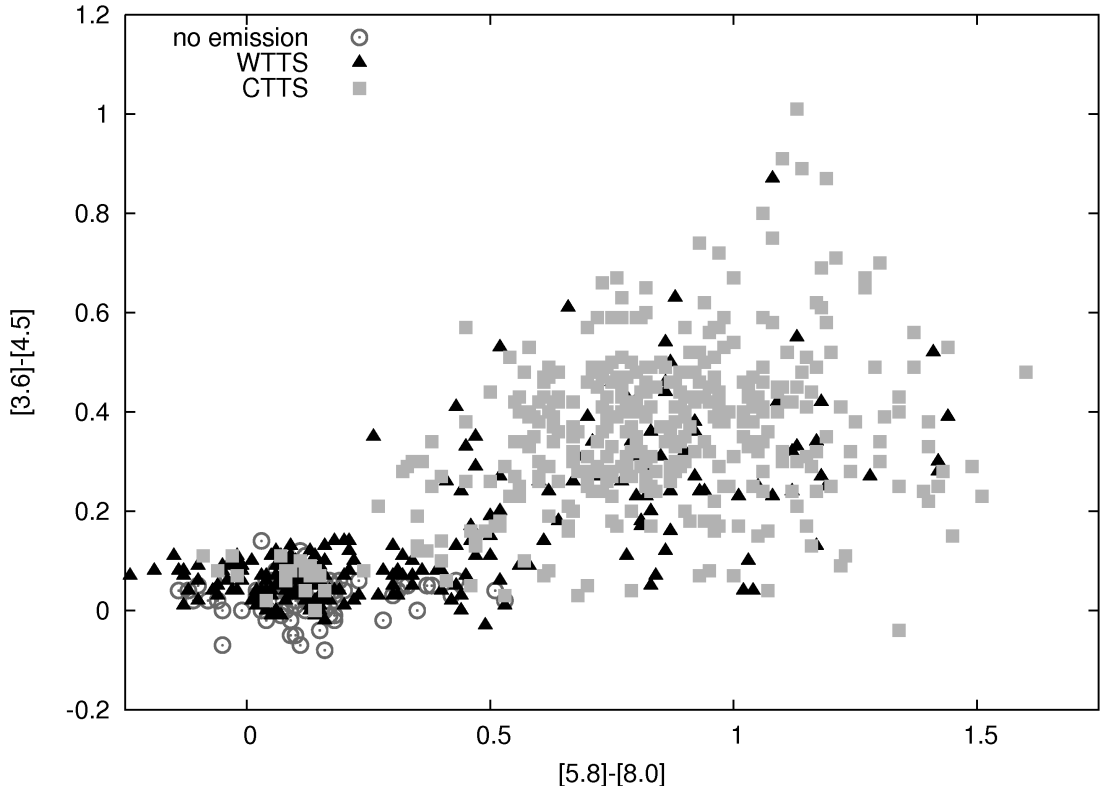}
\caption{IRAC color-color diagram of stars with measured RV. Stars with signs
of accretion (wide $H\alpha$ emission, CTTSs) detected in the spectra must have a disk,
and accordingly they do exhibit infrared excess (filled squares), occupying the approximate
domain of Class~II objects \citep{alle04}. Among stars showing only weak $H\alpha$ emission
(filled triangles) some still exhibit infrared excess, meaning the disk is yet presented
but accretion has already stopped. Most of the WTTS stars, however, have no signs of disks
and scatter around $(0,0)$. These Class III objects, exhibiting no intrinsic infrared excess,
share the area of the graph with field stars and older cluster members showing no signs
of $H\alpha$ emission (open circles).
\label{fig:irac_ccd}}
\end{figure}

Fig. \ref{fig:irac_ccd} shows that as expected, many weak-emission stars have no
infrared excesses and many wide-emission stars have infrared excesses, but
there are a significant number of counterexamples.  We suspect that many, if not
all, of the stars with significant [3.6]-[4.5] excesses are in fact accreting,
based on the strong correlation previously seen in K-L in Taurus (Hartigan \etal 1990),
but the strong nebular contamination prevents us from detecting the H$\alpha$ wings.
Other objects with only long-wavelength excesses may indeed not be accreting.
In addition, there are systems without any excesses which clearly are accreting.
In the cases of 0533477-045208 (F11\_ap43) or 0535343-060542 (F31\_ap158)
the IRAC colors are $[3.6-4.5]=0.06~/~0.07$, $[5.8-8.0]=0.41~/~0.13$,
respectively, both stars have high $S/N$ ratio spectra with very prominent,
wide \halp~ emission containing only negligible, easily identified nebular components.
In contrast, 0535513-061353  (F31\_ap208) and 0536222-055547 (S3\_ap217)
the IRAC colors show clear excess, $[3.6-4.5]=0.39~/~0.34$, $[5.8-8.0]=0.87~/~0.75$,
but the well defined spectrum shows only a weak nebular component with
slight asymmetry or very low level wings.

F11\_ap43 and F31\_ap158 appear to be examples of a small class of objects termed
``transitional disks'' (Calvet et al. 2002, 2005).  We have found $\sim35$ such 
stars in our ONC sample. These systems tend to have their
inner disks partially or almost totally cleared of small dust, as inferred by
the weakness of the near-infrared excess, yet have substantial outer disks.
TW Hya (Calvet et al. 2002, 2005) and DM Tau (Calvet et al. 2005) are examples
of systems which have essentially zero measurable excesses shortward
of $\sim 5 \mu$m but have strong excesses at longer wavelengths,
and {\em in addition} are still accreting gas onto the central star, producing
a broad H$\alpha$ emission profile.  
(For example, the 10-Myr-old accreting T Tauri star TW Hya has IRAC colors of
[3.6] - [4.5] ~ 0, [5.6] - [8] = 0.95; Hartmann et al. 2005.)
Megeath (personal communication) has confirmed the transitional disk
status of these objects, as his Spitzer photometric survey extended into the
24 $\mu$m band and some of these peculiar stars had usable (acceptable
S/N ratio) measurements exhibiting long wavelength excess.

How small dust ``disappears'' while gas
still accretes in the inner disk is not entirely clear.  One possibility
is that grains grow to such large sizes that the near-infrared opacity is
reduced; another possibility is that ``filtration'' occurs at the inner
edge of the optically-thick outer disk, moving particles of sizes responsible
for the near-infrared emission outward (Rice et al. 2006).  In any event,
the accreting transition disks we identify here in the ONC must be among
the very youngest such systems known, expanding the range of ages where
transitional disk behavior occurs.

\subsection{Binary Stars}

As the velocity dispersion in the ONC region is relatively small, identifying
spectroscopic binaries is important to explore intrinsic kinematic structure.
To ensure a consistent zero-point 
(see Section \ref{datared}) between the
multi-epoch observations discussed here, we included some stars of our first observing
run in the second set.
Also, to make sure that different fields of a given run have a common zero point,
we had some overlap between fields so that some stars were observed multiple times within
an observing run. 
Out of the 1215 targeted stars 1086 were observed only once, 122 twice,
and 7 three times. For the 129 stars observed multiple times, we compared
RV values and evaluated the difference ($\Delta$)
against the internal RV error ($\sigma$) of the correlation, identifying possible
($\Delta>1.5\sigma$) and likely ($\Delta>3\sigma$) binary stars. The \textit{CCF} column
of Table \ref{table:hecto_targets_mem} and \ref{table:hecto_targets_nonmem}
lists \textit{dv?} and \textit{dv} notes for these stars, respectively.

In case of a single RV measurement the cross correlation function still can
be used to identify double-lined spectroscopic binary stars.
By looking at each individual CCF we found several side lobes, secondary
peaks blended with the main peak (CCF note \textit{s}), and resolved
double peaks (\textit{d}). In case the main/secondary CCF peak heights
were not at least 50\% higher than any other local peaks due to noise,
or the side lobe did not raise 25\% higher than local peaks
and therefore the detection probability was not high,
we added the ``?'' sign in the notes to express the uncertainty.

\begin{figure}
\includegraphics[angle=270,scale=0.7]{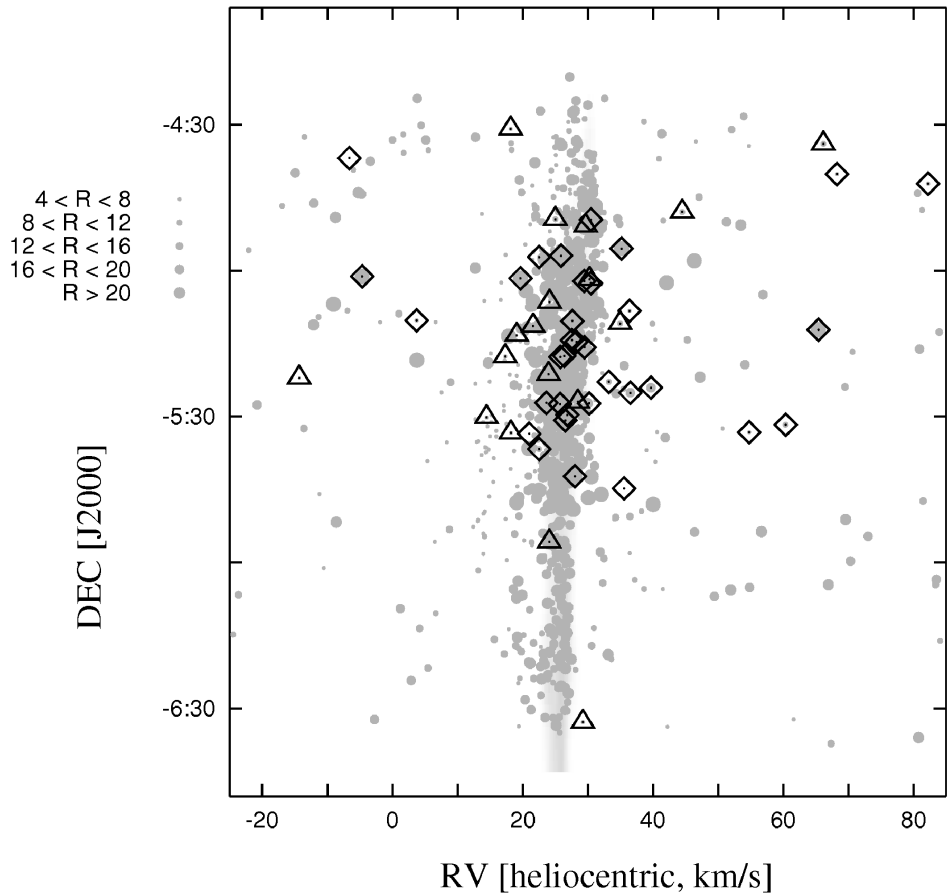}
\caption{Binary stars displayed on a spatial--velocity plot of the ONC region
(full projection along RA). The molecular gas is mostly buried by the
crowded stellar symbols (gray circles; size is correlated to R value like on
Fig. \ref{fig:rv_chanel}).
Binaries, identified by clear double peaks in the cross-correlation function or
by two different RV measured at two epochs, are shown as open diamonds.
(In case of more than one RV value the star is displayed according to the
average of the velocities.) Possible binaries, exhibiting asymmetric or 
side-lobed cross-correlation function, are displayed as open triangles.
\label{fig:full_ra_binary}}
\end{figure}

The location of these binary stars are shown in Figure \ref{fig:full_ra_binary}, displayed
as a full projection in RA, onto a RV--declination plane. The possible (more uncertain)
stars are shown with open triangles, the more likely candidates noted with open diamonds.
The velocity used for plotting is the average of all measured RV values in case we
had more than one measurement. A wider velocity range is applied to this plot
to accommodate all the binary stars, rendering the main stream cluster members and 
molecular gas into a narrow vertical feature. Counting all the possible (18)
and more likely ones (34), the total number of binaries is 52, or $\sim$4\% of the
entire sample.

Based on these single epoch observations we cannot make an estimate
of the spectroscopic binary fraction of the ONC.  However, our results may be consistent with
the low frequency of resolved multiple systems suggested for the ONC 
(see \citet{kohl06} and references therein).
In any event, our observation of position-velocity structure in the stellar
distribution, which strongly correlates with the molecular gas, indicates that
the overall dynamics are not strongly biased by binary motion, though our velocity
dispersions may be slightly inflated due to orbital motion.
Of course, there may be spectroscopic binaries present in our sample with much
smaller velocity shifts that are difficult to separate from the main population
(see \S 6.3).

\subsection{High Radial Velocity Stars}

In section \ref{velo} we showed that the RV distribution is relatively clean,
with very few outliers from the main peak (see insert of Fig. \ref{fig:rv_hist}).
Although most of these high velocities are just false detections (the CCF peak
found to be global is just marginally higher than any other local peak due to 
noise), some are real (a global CCF peak more than 50\% higher than other local peaks); these
stars have a note ``R'' in the \halp~ column of Table \ref{table:hecto_targets_nonmem}.
What makes these stars interesting is the possibility that they 
originated from a cluster by a decay of a triple or multiple young stellar system,
or by interaction of multiple systems \citep{gual04} and then were ejected at high velocities.
The very high stellar density of ONC makes it a favorable place to look for
such stars (one example may be the BN object, just north-west to
the Trapezium \citep{rodr05}). 
Three members of ONC were recently suggested by \citet{pove05} to be possible runaway
stars based on proper motion measurements performed on photographic plates, but
\citet{odel05} did not confirm these motions using 
proper motion measurements from Hubble Space Telescope images.
Only one of these three stars (JW 355) can be found in our sample as 
0535109-052246 (S2\_ap220). Unfortunately the spectrum is very noisy
so we have no additional radial velocity measurement.

Among our targets we have 21 stars with the ``R'' note, out of which
5 stars have velocities smaller than $-40$ \kms~ (more than $\sim20\sigma$ off the
mean cloud velocity) and 15 have RV values larger than 80 \kms.
Each of these stars were observed only once so the only hint for companions,
as an other possible reason of deviant RV, would be the CCF.
However, for all of them it only has a single, well defined peak. 

A relation to the cluster is apparent only in one case,
as star 0535503-044208 (F11\_ap146) shows
\halp~ emission (therefore listed in Table \ref{table:hecto_targets_mem}).
All the others exhibit only a nebular emission component superimposed on
the \halp~ absorption line. According to the stellar parameters
derived from the multi-template fitting, more than half of these
stars are K--M dwarfs. Assuming main sequence colors \citep{keny95},
based on the derived temperature, and comparing it to the observed
(J-H) value, we got extinction values of $A_{V}\simeq 3 - 4$ by
assuming a $E(J-H)=0.19~E(B-V)$ relation \citep{bess88}. This is a plausible
value in those areas these stars are located, farther from
dense regions.

The three most deviant RV stars are worth mentioning.
One object is blueshifted at $RV=-147$ \kms,
and exhibits only $A_{V}=1.3$. The low extinction is not surprising as this star
(0537111-055946 or F31\_ap162) is located at the very edge of our field,
which also means just off to the side of the Orion A molecular cloud.
The surface gravity suggests a giant, thus it is likely a background
object based on its brightness. This is in agreement with its location,
as we could have not observed it behind the denser parts of the cloud.

The two highly redshifted stars (0536592-050029 or F11\_ap183
at $RV=232.5$ \kms; and 0537005-050931 or F22\_ap138
at $RV=438.7$ \kms) have almost the same RA value as F31\_ap162, so
those are also just off the cloud.  The former star seems to have a
low surface gravity, and therefore it is likely a background
object as well, at a brightness of $J=11.87$.
However, there seems to be a 
contradiction with its J magnitude, as the $(J-H)=1.09$ color
suggests a very high extinction adopting the intrinsic
colors of $(J-H)_0=0.66$ from \citet{alon99}, based on the
stellar parameters derived form template fitting.
Another problem with the interpretation as a background object
is that the $[Fe/H]=+0.5$ templates used for the analysis might be not adequate.
Re-running the cross-correlation on a grid of metallicities
(with a fixed rotational velocity of 0) it turns out a $[Fe/H]=-0.5$
gives a better match (with $T_{eff}=4500$ and $log(g)=1.5$), but 
the observed and intrinsic colors are still in contradiction with
the expected low extinction. Thus, additional spectroscopy is 
required to say more about this object. 

Exploring metallicity by further template matching turned out
to be crucial for the highest radial velocity star, F22\_ap138, as 
for temperature we hit the edge of parameter space
while performing cross-correlation of the original grid.
Fixing the rotational velocity at
0 and exploring a metallicity--temperature--gravity grid
we found the best match to be a $[Fe/H]=-1.5$
spectrum with $T_{eff}=4250$ and $log(g)=4.0$.
This results a plausible extinction of $A_V=1.1$.
The low metallicity template suggests that this is not a young
star, therefore very likely we witness a fast moving background
member of an older population.

\section{Discussion}

\subsection{Cluster in Formation}

The significant position-velocity substructure seen in both the 
stars and the molecular gas clearly demonstrates that the ONC is not
dynamically relaxed; qualitatively, it appears to be more consistent
with dynamical models of cluster formation such as those of Hartmann
\& Burkert (2007) and Bate, Bonnell, \& Bromm (2003).  
The curvature of the $^{13}$CO emission seen in the northern region of
the position-velocity plot (Figure 4), 
with a possible corresponding southern feature,
is suggestive of gravitational acceleration of material toward the cluster center,
where the gravitational potential well should be deepest.
Gravitational acceleration is also consistent with the evidence for the
largest velocity distribution in the gas located near the center. 
Many stars exhibit the same position-velocity distribution as the $^{13}$CO,
which suggests that these stars are mostly following the motion of the
dense gas within which they formed.
If these motions were primarily the result of stellar energy input through winds or 
photoionization, it is not clear that the gas and many stars should show the same kinematics,
as such energy input could easily blow the gas away from the stars - some possible
examples of this are discussed in the following section.  

The observed correlation of stars and dense gas in space and in velocity indicates that
the inner regions cannot have experienced more than about one crossing or collapse
time; otherwise the velocity structure would be erased as the infalling gas shocks
and dissipates its kinetic energy, with the stars passing through the shocked gas.
Overall, the spatial coherence in the $^{13}$CO gas kinematics, the evidence
that many younger stars follow this motion, and the strongly filamentary
structure of the stellar distribution outside the main core of the ONC
suggest that the ONC still exhibits features traceable to its initial conditions of formation.

\subsection{Subgroups and stellar energy input}

\begin{figure}
\includegraphics[angle=270,scale=0.8]{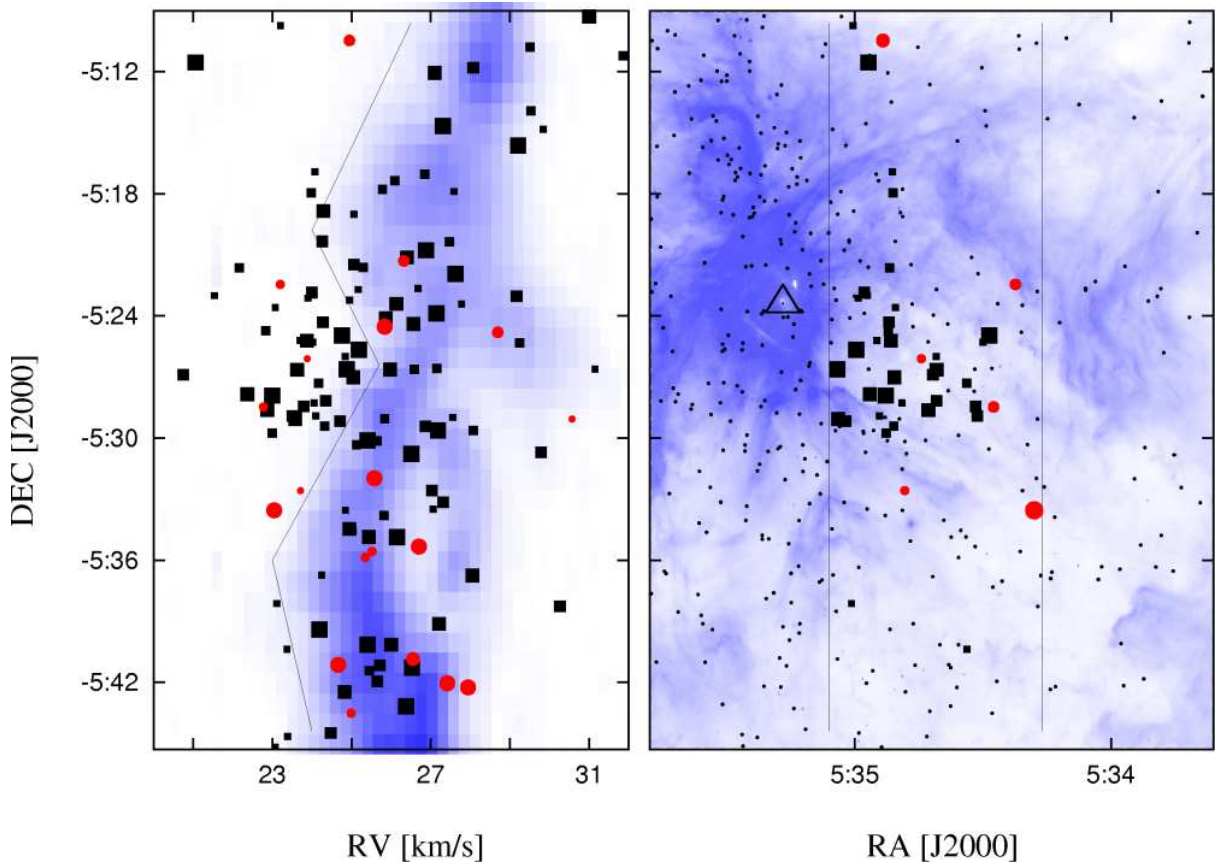}
\caption{In the left panel, we show
a zoomed-in velocity--declination channel map similar to the bottom
left panel of Fig. \ref{fig:ra_chanel}.  A group of stars 
in the range  $-5^{d}~20^{m}~>DEC>~-5^{d}~30^{m}$ is slightly but clearly offset
from the molecular gas by about 1-2 \kms.  Most of these stars also show infrared excess
(filled squares); only a few do not (open circles), and so are clearly
members.  We defined a velocity envelope for the $^{13}$CO emission (thin curve), and 
plot the positions of the stars blueward of this envelope in the right panel.
These ``off-cloud'' stars tend to lie in a relatively evacuated region 
west--south-west of Trapezium (noted as an open triangle), as indicated
by the low dust emission seen in the $8 \mu$m IRAC map (shaded area).
All other stars with measured velocities and $R>4$
are shown as black dots, and two vertical lines represent the RA limits of the channel map
on the left.  The morphology of the region suggests that molecular gas has recently
been evacuated by outflows from the Trapezium region (see text).
\label{fig:nose_proj}}
\end{figure}

While there is a strong general spatial and kinematic correlation between stars and
dense gas, there are exceptions.  One especially clear case can be seen
on the bottom left panel of Fig. \ref{fig:ra_chanel} at
$DEC\simeq-5\degr27\arcmin$: a tight, linear string of
stars  sticking out of the main cloud as a ``nose'' extending between
$RV=21$ and 25 \kms.
The left-hand panel of \ref{fig:nose_proj} shows a declination-velocity plot zoomed
into this region, in an RA range shifted slightly westward of the Trapezium
region (vertical lines in the right-hand, RA-DEC panel of Fig. \ref{fig:nose_proj}).  Clearly there is a small group
of stars blueshifted by about 1-2 \kms~ from the gas in the declination range
$-5^{d}~20^{m}~>DEC>~-5^{d}~30^{m}$.  Selecting ``off-cloud'' stars by means
of an velocity envelope for the gas (thin curve in left panel) results in demonstrating
that these stars are spatially concentrated in a region south-west
of the Trapezium which is relatively evacuated, as indicated by the weak $8 \mu$m emission
seen in the IRAC map (right panel, grayscale).  This region is also known for a large number of
Herbig-Haro objects shaped like bow-shocks pointing back toward the Trapezium region
(see description in \citet{odel01}), consistent with the idea that outflows
from the region of most current star formation is blowing out material. 

We therefore suggest that the molecular gas in this region originally extended
to more negative velocities, but that outflows have cleared away the gas,
making an ``indentation'' in the position-velocity plot at $\sim -5^{d}~26^{m}$,
RV $\sim 24$ \kms, and leaving the recently-formed stars behind.

There are other interesting small scale structures, like
the tight, dense sub-cluster group seen in the middle of the
second panel of Fig.  \ref{fig:dec_chanel} (upper row, 2nd
panel from left to right, also somewhat apparent on Fig.
\ref{fig:ra_chanel} upper row, rightmost panel).
These can be remnants of small building blocks, 
traces of density variations in the primordial cloud.

\subsection{Other Populations?}
\label{foreground}

\begin{figure}
\includegraphics[angle=270,scale=0.65]{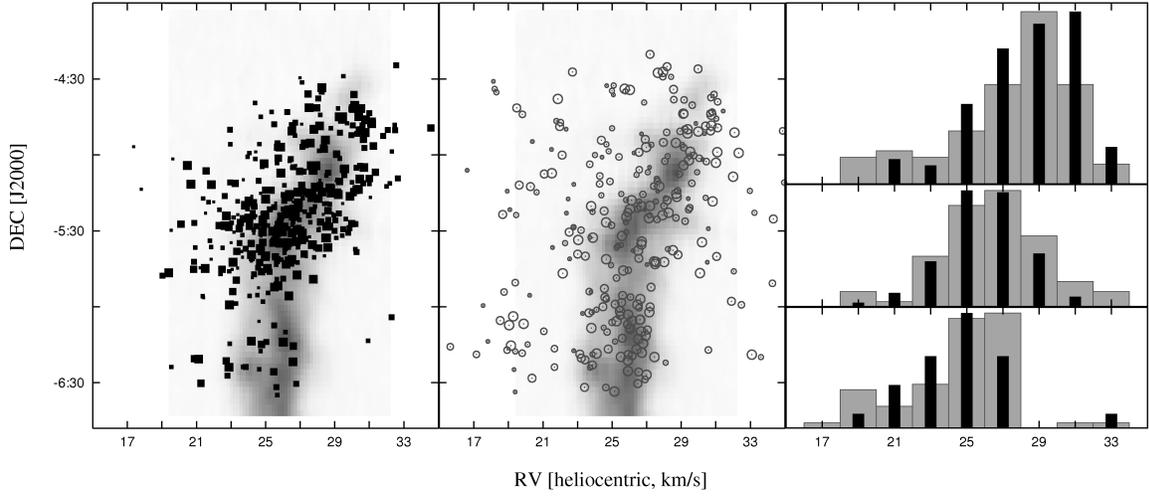}
\caption{Velocity--spatial plot, full projection in RA, of stars with
(filled squares, left panel) and without (open circles, middle panel)
infrared excess. On the right, we show histograms for three declination
bands, which are indicated by the locations of the histograms. Note, that
for the northern section ($DEC>-5.2$ deg) the mean velocity of the gas is
redshifted, as it is seen in the displaced peak of the distribuiton
as well. The middle ($-5.2>DEC>-6.0$) and southern ($DEC<-6.0$) sections
have similar mean velocities, although there is a strong RV gradient
in the mid-declination range. On these histograms the
gray shaded columns represent the distribution of the non-excess stars
(middle panel), the black bars represent the excess stars (left panel).
Note the group of
stars around $DEC=-6^{d}~15^{m}$ and $RV=19$ \kms, which is only 
apparent on the middle panel (non-excess stars) and could be an
older foreground population. This group is responsible for the 
small local peak at 19 \kms~ in the RV histogram of Fig. \ref{fig:rv_hist}.
Also note that at $DEC<-6$ there are less IR excess sources on cloud, which
is a selection effect (see text for details). 
\label{fig:irac_wtts}}
\end{figure}

Based on Fig. \ref{fig:irac_ccd}, we divided our sample into 
\textit{IR-excess sources}, with $([3.6]-[4.5]) > 0.2$ and
$([5.8]-[8.0]) > 0.5$; and \textit{non-IR excess sources} with
$([3.6]-[4.5]) < 0.2$ and $([5.8]-[8.0]) < 0.5$.
This division of sources is shown on the velocity-declination map
with full projection along RA in Figure \ref{fig:irac_wtts}:
the left panel for excess and the middle one for non-excess stars.

Looking at the left panel
in Figure \ref{fig:irac_wtts}, southward
of $-6\degr$ there are very few CTTS.
This abrupt change is an unfortunate selection effect.
The targets selected for our second observing run (S fields)
had a southern declination limit of $-6\degr$, exactly where the sudden
change of excess/non-excess ratio happens. And since targets for the first observing
run (F fields) were selected from the 2MASS color--magnitude
diagram, our selection was biased against stars with IR excesses
because of the $(H-K)<0.5$ criterion.

While many of the non-excess stars (middle panel) follow the general distribution of gas
and excess stars, a modest number clearly exhibit a broader distribution in radial
velocity, with more blue- than redshifted stars. This can be seen on the right panel,
where we present RV histograms of excess (solid bars) and non-excess (gray-shaded
boxes) stars, for three declination ranges (north, with $DEC>-5.2$; mid, with 
$-5.2>DEC>-6.0$; and south, $DEC<-6.0$ - as represented by the location of the
histograms as well). We divided this way because
RV bahavior of the gas is very different in these regions:
the northern velocity is significantly higher than the mean velocity in the
south, and the middle section exhibits a strong RV gradient.
Regardless, the velocity distribution of non-excess stars is wider because
of a secondary peak at lower velocities.

One possibility is that the lower-velocity stars have been ejected from the main
cluster.  Close encounters in multiple star systems might result in ejecting
stars while stripping their disks; the red-shifted stars might then eventually plunge
into the dense, opaque regions behind, leaving more optically-visible stars with
blueshifts.  However, this seems unlikely because
it requires a surprisingly large fraction of stars to be
ejected, we do not detect a larger density of such objects nearer
the center of the ONC, where the high stellar density would be
more favorable to ejection. In addition, south of $-6^{\circ}$ the velocity distribution
is clearly bimodal, with a very tight velocity spread for the on-cloud stars; it
is not clear why ejection would lead to such distinct structures.

Another possibility is that we are detecting a foreground, older population.
Stars in the foreground Orion 1a association 
(Brown \etal 1994) may have a heliocentric radial velocity
near $\sim 20$ \kms, as suggested by measurements of the 25 Ori group
\citet{bric07}.  If this is the case, it would mean that the Orion 1a
association is much more spread spatially towards the ONC (in the foreground)
than has previously been realized. 

\begin{figure}
\includegraphics[angle=270,scale=0.8]{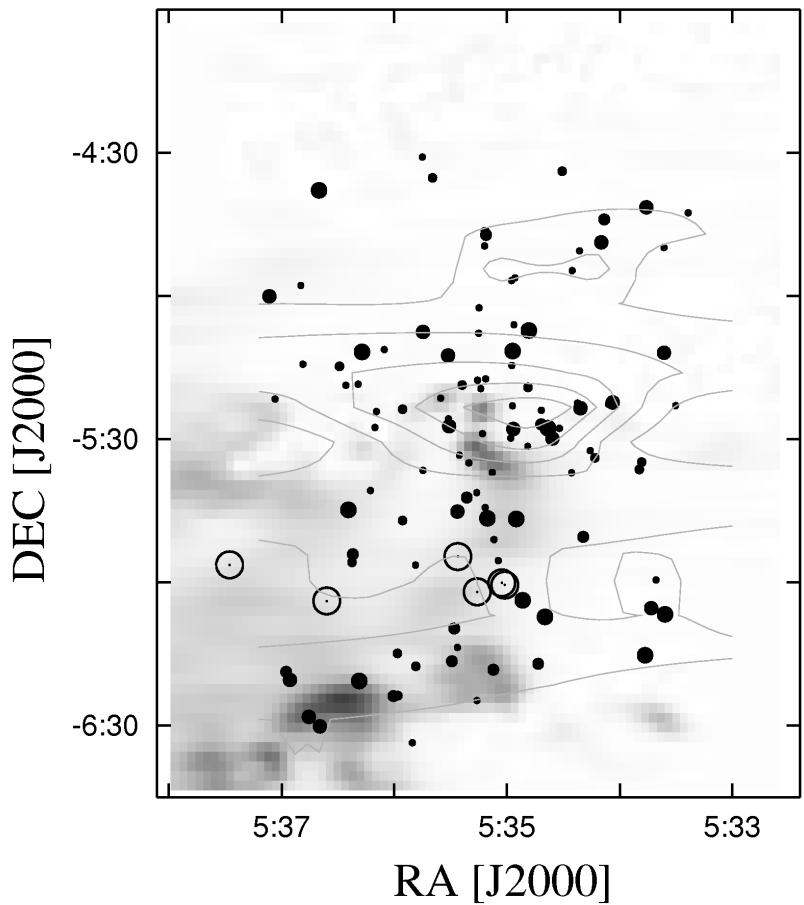}
\caption{Spatial distribution of foreground stars (filled circles),
with significantly blue-shifted velocities respect to the molecular
cloud (shaded areas). The stars are displayed between 
$16< RV < 22$ \kms, and since there is no gas at this heliocentric velocity range,
we plot the most blueshifted RA-DEC channel of the gas at $RV=23$ \kms. Contours of stellar
density are also plotted (with levels of 1 star per 0.2$\times$0.2 degree area)
Open circles note the OB stars arund $DEC\simeq-6\degr$ 
(HD 37303, 37209, 37043, 37025, 36960 and 36959).
\label{fig:foreground}}
\end{figure}

A third possibility is that stellar energy input blew out material,
forming a small proportion of stars in a bubble wall moving toward
us.  To explore this idea further,
we plot only the stars with RV values less than 22 \kms
in the RA--DEC plot of Figure \ref{fig:foreground}.
Since there is no gas here, we display the most blueshifted
1 \kms~ wide channel of the gas at 23 \kms~ heliocentric velocity.
The northern foreground stars seems to be separated from the
older southern group, by an almost empty gap running
east--west at $-6\degr00\arcmin$. However,
this might be the result of a combination of our bias against
selecting infrared-excess stars mentioned above and accidental fiber
coverage. Still, there is some structural similarity between
the distribution of these southern foreground stars and the
gas ``behind'', redshifted to it. Note how these
southern stars are spread on top of two-three denser gas clumps.
If this structure is interpreted as a local ``bubble'',
this could have been created at the very beginning of star formation
by the early emergence of some OB stars in the region.
These stars than could have blown away most of the gas from us,
some smaller amount towards us,
and so other stars could have formed at this early stage
in this perturbed southern region. 

This picture would require some OB stars in the questioned region, and the
list of \citet{brow94} contains six of them
interestingly close to $-6$ degree declination
(see Fig. \ref{fig:foreground}).
However, obviously none of these is in close
vicinity of the gas now, as there are no apparent
HII regions at this part of the molecular cloud.
An additional problem is that one would expect such a population
to be somewhat older than the main cloud, which is not obvious in the J vs. J-H
diagram.  This is not a particular problem for the foreground population hypothesis,
as the Orion 1a stars will appear higher in the color-magnitude diagram at a given
age than the ONC.

In conclusion, the reasons for the additional velocity spread of the non-excess
stars are not clear.  Further monitoring for radial velocity variations as well
as more detailed studies of the stellar properties are needed to determine which
of the above possibilities are correct.

\subsection{Comparison with Other Observations and Models}

If the ONC is not dynamically relaxed, why did HH98
find a fairly good fit to a King cluster model?  (Note that
\citet{scal05} point out difficulties with physically
interpreting the ONC in this way). The central region 
of radius $\sim 0.5$~pc does appear to exhibit a smooth distribution, and thus
may be closer to having relaxed, though it is somewhat elongated.
On larger scales, the elongation is sufficiently large that azimuthal
averaging is not appropriate.  To take an extreme example of
what this averaging can do, imagine a cylindrical distribution
of material in the plane of the sky with uniform surface density and width $W$.
Azimuthal averaging of this distribution results in an average surface density
$\Sigma \sim$~constant for $r < W$ and $\Sigma \propto r^{-1}$ for $r \gg W$.
Interpreting this surface density in terms of a spherically-symmetric density
distribution results in 
$\rho \sim$~constant for $r < W$ and $\rho \propto r^{-2}$ for $r \gg W$,
which is roughly consistent with a King model.  
The ONC is not as extreme in structure as this simple example, 
but it illustrates the potentially misleading nature of azimuthal averaging
of an intrinsically highly elongated structure.

Comparing azimuthally averaged stellar densities and
global velocity dispersions to N-body calculations,
\citet{krou00} was unable to distinguish whether the ONC is in equilibrium,
or is collapsing, or is expanding; a similar result was obtained
\citet{scal05}.  This demonstrates the importance of
avoiding azimuthal averaging and obtaining detailed kinematic observations 
of gas and stars in developing an understanding of the dynamical
state of the region.

Tan, Krumholz, \& McKee (2006; TKM06) have argued for a picture in which 
rich star clusters take several dynamical times to form, are quasi-equilibrium 
structures during formation, and thus initial conditions are not very important.
In particular, TKM06 discuss the ONC and argue that it is several crossing times
old, of order 3 Myr, larger than assumed here.  Here we discuss why
we arrive at different conclusions.

Following \citet{scal02}, TKM06 argue that the smoothness 
of the spatial distribution of the stars in the ONC argues for long formation 
times, which allow clumps of stars to disperse.  As pointed out above, the azimuthal
averaging done by \cite{hill98} was somewhat misleading,
helping to smooth out filamentary structure.  In addition, the gas kinematics,
and to a lesser extent the stellar kinematics, show substructure which is less
evident in the spatial distribution alone.

TKM06 also use the analysis of stellar ages in the $0.4 - 6 \msun$
mass range by \citet{pall99} to argue for a large age spread in the ONC.
Although they note that \citet{hart03} 
pointed out problems of contamination by non-members, TKM06 do not take
sufficient notice of the mass-dependence of isochrones.  
As was clear from the original work by \citet{hill97},
and is evident in Figure 4 of \citet{pall00}, standard isochrones
imply that stars of 1 to 5 $\msun$ in the ONC have systematically older ages by one 
to a few Myr than both the lower mass stars and the higher mass stars.
This effect is seen in virtually every star-forming region, at least
comparing the intermediate mass stars to the lower mass $< 1 \msun$
stars (see HR diagrams in \citet{pall00}).  In other words, the
``tail'' of older stars in the distribution of stellar ages found by Palla \&
Stahler (1999, 2000) is strongly populated by a {\em different} mass range
(intermediate mass stars) than the peak of the age distribution (low mass stars).

As pointed out by \citet{hart03}, if one takes these age determinations at face 
value, it means that the ONC existed for a few Myr as a cluster
forming mostly intermediate mass stars, with very few low mass stars (or high-mass stars).  
In other words, the ONC (and other regions) begin their first few Myr of existence with 
extremely non-standard stellar initial mass functions (IMFs).   There is no observational
evidence for any young clusters dominated by intermediate-mass stars by number.
Therefore, there must be something wrong with the isochrones.

Pre-main sequence ages are contraction ages; one must know the starting radius
of the star to determine the age.  TKM06 argue that corrections for starting
radii (the ``birthline'' position) are only important for stars with ages $< 1$~Myr.
However, \citet{hart03} showed that this is not necessarily true for intermediate-mass
stars; changing the accretion rates at which stars form over plausible ranges can
have very big effects on the birthline for $\gtrsim 1 \msun$ stars.  Thus, 
the apparent ages of the intermediate-mass stars can be
systematically overestimated due to incorrect birthline assumptions; this allows
the ONC to maintain a roughly typical IMF over its formation.
With this correction, the fraction of stars older than 2 Myr decreases dramatically.

\section{Summary}

We have carried out a spectroscopic survey of 1215 stars
located in the northern end of Orion A molecular cloud,
covering the ONC and its vicinity. The obtained radial
velocities show a well defined spatial and kinematical
structure of the stellar component, which is very similar
to the one seen in the molecular gas. Comparing our
observational results to the model of \citet{hart07},
we draw the following picture of the ONC region:

On large scales the gas (and stars) exhibit a velocity gradient due to
rotation or shear running north-south.  The curvature seen most
clearly in the northern arm of the gas in the position-velocity
diagrams suggests gravitational acceleration towards the cluster center. 
We further conjecture that the concentration of gas 
(and stars) south of the Trapezium region is somewhat in front
and is also falling in towards the center, explaining its higher radial
velocities; the Orion Bar photodissociation region appears to reside just
above, or perhaps at the upper edge, of this moving clump.  The
southern part of the filament may also be falling in, although the
motions are much less organized than is apparent in the northern arm. 
Finally, it may be possible that the process of blowing out the near side of the
cloud, in the south, resulted in accelerating and compressing gas which formed a small
population of stars with velocities blueshifted by a few \kms; or, 
these stars may be foreground objects in a different kinematic system
possibly associated with Orion OB 1a.

The high degree agreement between the structure of the
gaseous and stellar component suggest the region is very
young, only $\sim1$ crossing time old, otherwise
gravitational interaction should have been smoothed
the fine structure still clearly visible in our data.
Note that the observational errors for the stellar radial velocities
(ranging from 0.5 - 1.5 kms) are larger than the observational errors
in the radio measurements.  In addition, with only two epochs (at most)
we cannot correct for the effects of even small binary motion. For
these reasons it is not yet possible to make fine distinctions between
stellar and gas kinematics.

The spectra in the \halp~ order provide not only RV measurements but
allow the possibility of classifying many stars
according to their \halp~ emission. 
With the aid of IRAC photometry we were able to
confirm the presence of disk around many CTTS 
opening/loss of a disk around most WTTS.
To the south we see another older but foreground population,
which still might be not completely independent but connected to the region,
suggested by continuous streams of stars to the south and north
of this group and by the similarity in the spatial structure 
of the ``background'' gas and these blueshifted stars.

In a future communication we will provide further improvements
in the detection of spectroscopic binaries and provide additional
checks on membership using the 6708 \AA~ Li line.

\acknowledgments
We would like to thank for the MMT observers and
instrument support team for all their efforts to make
Hectochelle work and for providing continous help in the 
preparation of the observations and in data acquisition:
Perry Berlind, Mike Calkins, Maureen Conroy, Dan Fabricant and
John Roll. Lori Allen's comments were very useful 
in making the text comprehensive.
This work was supported in part by NASA grant NNG06GJ32G.

\clearpage

\clearpage

\begin{deluxetable}{cccccc}
\tabletypesize{\scriptsize}
\rotate
\tablecaption{Summary of spectroscopic observations \label{table:obs}}
\tablewidth{0pt}
\tablehead{
\colhead{Date} & \colhead{ID} & \colhead{field center} & \colhead{exposures}
& \colhead{binning mode} & \colhead{number of targets}\\ 
}
\startdata
2004 Dec 01	& F11	&    $5^{h}~35^{m}~19^{s}~~ -04\degr~49\arcmin~27\arcsec$ &  $3\times20$ min  & $1\times1$  & 224\\
2004 Dec 01	& F21	&    $5^{h}~35^{m}~13^{s}~~ -05\degr~21\arcmin~14\arcsec$ &  $3\times20$ min  &  $1\times1$  & 210 \\
2004 Dec 01	& F22	&   $5^{h}~35^{m}~13^{s}~~ -05\degr~21\arcmin~14\arcsec$ &  $3\times20$ min  &  $1\times1$  & 219\\
2004 Dec 01	& F31	&    $5^{h}~35^{m}~26^{s}~~ -06\degr~07\arcmin~30\arcsec$ &  $3\times20$ min  &  $1\times1$  & 213 \\
2005 Nov 15	& S3	&    $5^{h}~35^{m}~13^{s}~~ -05\degr~33\arcmin~10\arcsec$ &  $3\times15$ min  &  $2\times2$  & 161\\
2005 Nov 15	& S2	&   $5^{h}~35^{m}~11^{s}~~ -05\degr~07\arcmin~28\arcsec$ &  $3\times15$ min  &  $2\times2$  & 154\\
2005 Nov 15	& S1	&   $5^{h}~35^{m}~21^{s}~~ -05\degr~18\arcmin~50\arcsec$ &  $3\times15$ min  &  $2\times2$  & 170\\
\enddata
\end{deluxetable}

\begin{deluxetable}{cccccccccccccccc}
\tabletypesize{\scriptsize}
\rotate
\tablecaption{Members \label{table:hecto_targets_mem}}
\tablewidth{0pt}
\tablehead{
\colhead{2MASS\_id} & \colhead{ID} & \colhead{J} & \colhead{$(J-H)$} & \colhead{$(H-K)$} 
& \colhead{$[3.6-4.5]$} & \colhead{$[5.8-8.0]$} & \colhead{$V_{rad}$} & \colhead{$\Delta V_{rad}$} & \colhead{R} & \colhead{S} & \colhead{$EW_{c}$} & \colhead{$FW_{10\%}$} & \colhead{$H\alpha$} & \colhead{CCF} &\colhead{NOB} \\
}
\startdata
0533179-052138	& F22-ap25	& 13.18	& 0.87	& 0.24	& 0.00	& 0.00	& 25.09		& 6.38	& 4.4	& 0.39	& 11.2	& 239	& W	& w	& 1\\
0533204-051123	& F21-ap89	& 12.73	& 1.07	& 0.29	& 0.00	& 0.00	& 24.28		& 0.53	& 13.6	& 0.58	& 5.2	& 169	& W	& c	& 1\\
0533225-053240	& F22-ap20	& 13.02	& 0.98	& 0.23	& 0.00	& 0.00	& 23.89		& 0.35	& 22.4	& 0.71	& 5.3	& 134	& W	& c	& 1\\
0533233-052153	& F22-ap27	& 12.90	& 0.87	& 0.24	& 0.00	& 0.00	& 47.20		& 0.35	& 23.4	& 0.74	& 0.9	& 179	& W	& c	& 1\\
0533234-044234	& F11-ap62	& 13.38	& 0.94	& 0.34	& 0.00	& 0.00	& 21.48		& 4.60	& 5.0	& 0.23	& 3.7	& 152	& W	& w,l	& 1\\
0533256-052354	& F21-ap61	& 13.32	& 0.95	& 0.21	& 0.02	& --	& 69.48		& 0.49	& 15.1	& 0.64	& 2.4	& 178	& W	& c	& 1\\
0533263-051640	& F22-ap37	& 13.34	& 1.25	& 0.33	& 0.00	& 0.44	& 70.65		& 0.64	& 10.7	& 0.54	& 5.2	& 189	& NSW	& c	& 1\\
0533285-051726	& F22-ap40	& 11.92	& 1.12	& 0.34	& 0.04	& 0.79	& 25.01		& 0.32	& 28.7	& 0.85	& 23.1	& 390	& C	& c	& 1\\
0533287-052610	& F21-ap69	& 13.23	& 1.08	& 0.29	& 0.02	& 0.12	& 29.23		& 0.82	& 12.1	& 0.53	& 6.6	& 116	& NSW	& c	& 1\\
0533293-050749	& F22-ap49	& 13.29	& 1.20	& 0.32	& 0.03	& 0.00	& 26.42		& 2.32	& 10.1	& 0.62	& 4.0	& 184	& W-	& c	& 1\\
0533301-052257	& F22-ap30	& 13.03	& 0.79	& 0.21	& 0.02	& 0.00	& 22.36		& 2.34	& 6.8	& 0.48	& 2.3	& 153	& W	& c	& 1\\
0533302-060409	& F31-ap51	& 12.26	& 0.89	& 0.22	& 0.00	& 0.14	& 32.28		& 3.01	& 14.8	& 0.81	& 7.0	& 448	& C	& w	& 1\\
0533307-051352	& F21-ap86	& 13.16	& 0.89	& 0.21	& 0.24	& 0.44	& 25.00		& 0.58	& 18.5	& 0.70	& 4.2	& 146	& NSW	& c	& 1\\
0533307-051351	& S2-ap26	& 13.16	& 0.89	& 0.21	& 0.24	& 0.44	& 25.32		& 0.63	& 16.9	& 0.72	& 2.0	& 164	& W-	& c	& 1\\
0533307-050813	& F21-ap94	& 13.02	& 1.11	& 0.31	& 0.01	& 0.02	& -6.91		& 0.73	& 10.4	& 0.60	& 0.0	& 0	& W	& c	& 1\\
0533310-060605	& F31-ap57	& 13.07	& 0.91	& 0.21	& 0.05	& 0.13	& 25.00		& 0.75	& 15.9	& 0.71	& 2.4	& 142	& W+	& c	& 1\\
0533312-052957	& F22-ap13	& 13.45	& 1.01	& 0.32	& 0.23	& 1.51	& 28.98		& 0.44	& 14.9	& 0.57	& 33.0	& 229	& C	& c	& 1\\
0533314-060954	& F31-ap52	& 12.75	& 1.08	& 0.36	& 0.34	& 0.94	& 26.16		& 0.88	& 11.2	& 0.51	& 52.1	& 293	& C	& c	& 1\\
0533329-055909	& F31-ap66	& 12.10	& 0.75	& 0.23	& 0.01	& 0.14	& 32.48		& 0.57	& 14.1	& 0.67	& 0.0	& 0	& AE	& c	& 1\\
0533330-051155	& S1-ap75	& 12.57	& 1.22	& 0.42	& 0.25	& 0.98	& 28.16		& 0.39	& 14.7	& 0.62	& 8.5	& 335	& C	& c	& 2\\
0533339-053326	& F22-ap14	& 12.21	& 1.25	& 0.40	& 0.26	& 0.67	& 27.11		& 0.40	& 21.2	& 0.76	& 4.9	& 181	& W	& c	& 1\\
0533343-061352	& F31-ap48	& 12.72	& 0.94	& 0.20	& 0.00	& 0.00	& 26.03		& 0.35	& 24.2	& 0.79	& 3.6	& 383	& C	& c	& 1\\
0533344-051417	& S1-ap78	& 12.48	& 1.12	& 0.38	& 0.47	& 1.04	& 27.12		& 0.60	& 17.8	& 0.72	& 4.1	& 308	& CW	& c	& 1\\
0533357-050923	& F21-ap92	& 12.50	& 1.09	& 0.28	& 0.06	& 0.08	& 27.59		& 6.19	& 8.1	& 0.51	& 4.9	& 354	& CW	& w	& 1\\
0533358-050132	& S2-ap44	& 11.72	& 1.34	& 0.51	& 0.43	& 0.93	& 24.41		& 0.81	& 10.7	& 0.58	& 39.1	& 473	& C	& c	& 1\\
\enddata
\tablecomments{Hectochelle targets in ONC found to be members based on measured RV value
or by detected \halp~ emission. The criteria of beeing RV member is to have at least one
velocity measurement within 4$\sigma$ of the cluster mean velocity:
13.7 \kms~ $< ~V_{helio} <$ 38.5 \kms~. RV members listed only with $R>2$, while stars
with \halp~ emission are included regardless of R value, but with no velocity displayed.
-- for the full table see the electronic version of the Jurnal or contact the authors)\\
~~~
\textbf{2MASS id} --- 2MASS identification number 
(truncated RA and DEC coordiantes as: HHMMSSS+DDMMSS);~~
\textbf{ID} --- internal identification number, specifying the field
(see Table \ref{table:obs}) and aperture of observation. The letter \textit{F} 
means the first run, and therefore 2MASS based selection, the letter \textit{S}
notes IRAC based selction and observation taken during the 2nd run;~~
\textbf{J} --- 2MASS J magnitude;~~
\textbf{(J-H)} --- 2MASS $(J-H)$ color index;~~ 
\textbf{(H-K)} --- 2MASS $(H-K)$ color index;~~
\textbf{$3.6-4.5$} ---  IRAC short wavelength color index;~~
\textbf{$5.8-8.0$} ---  IRAC long wavelength color index;~~
\textbf{$V_{rad}$} --- measured heliocentric radial velocity, in \kms;~~
\textbf{$V_{rad}$} --- \textit{xcsao} error estimate for $V_{rad}$, in \kms;~~
\textbf{R} --- R value of cross correlation (see text for details);~~
\textbf{S} --- height of the CCF peak;~~
\textbf{$EW_c$} --- absolute value of equivalent width, measured on the corrected
\halp~ profile (see text for details);~~
\textbf{$FW_{10\%}$} --- full width of \halp~ profile at 10\% level of the corrected
maximum (see text for details);\\
~~~
\textbf{H$\alpha$} ---  notes on the \halp~ emission profile. See text for
more details on the CTTS/WTTS classification:~~
 \textbf{C} -- CTTS;~~
 \textbf{W} -- WTTS;~~
 \textbf{CW} -- rather CTTS, but could be WTTS;~~
 \textbf{WC} -- rather WTTS, but could be CTTS;~~
 \textbf{W+} -- assymetry in line profile/wing: likley WTTS;~~
 \textbf{W-} -- assymetry in wings, but wings at low intensity: could be WTTS;~~
 \textbf{D} --- resolved or unresolved double gaussian profile, 
  no excess H-alpha emission (likley non-TTS);~~
 \textbf{CD} --- resolved or unresolved double gaussian profile, likely CTTS;~~
 \textbf{WD} --- resolved or unresolved double gaussian profile, likely WTTS;~~
 \textbf{R} --- obviously shifted H-alpha absorption : high RV stars;~~
 \textbf{X} --- very wide H-alpha absorption with nebular emissions;~~
 \textbf{CX} --- CTTS, with strange emission profiles;~~
 \textbf{AE} --- stellar absorption and nebular emission together;~~
 \textbf{NS} --- very strong nebular component component, no wings, no asymmetry;~~
 \textbf{NSC} --- like NS, but if intensity logscaled wide wings/assymetry is visible
 and suggest CTTS emission signature under the strong nebular component;~~
 \textbf{NSW} --- like NS, but if intensity logscaled narrow wings/assymetry is visible
 and suggest WTTS emission signature under the strong nebular component;~~
 \textbf{SAT} --- saturated, or neighbour saturated;\\
~~~
\textbf{CCF} --- notes on the cross-correlation function:
 \textbf{w} -- wide peak, results large errors in RV;
 \textbf{l} -- very low peak, but still sticks out from surrounding;
 \textbf{u} -- almost undefined, very low/wide peak;
 \textbf{n} -- very noisy (several local peaks, almost as high as the one picked);
 \textbf{s} -- side lobed, could be a spectroscopic binary;
 \textbf{d} -- double peak, spectroscopic binary;
 \textbf{dv} -- RV measured more than once, found different velocities, could be binary;
 \textbf{r} -- template spectra with highest peak in cross-correlation resulted non-realistic velocity,
so the template resulting in highest R value was used instead to determine velocity;
 \textbf{c} -- clear, well-isolated peak;
 \textbf{?} -- uncertainty in assigning the given note;\\
~~~
\textbf{NOB} --- number of observations}
\end{deluxetable}

\begin{deluxetable}{cccccccccccccccc}
\tabletypesize{\scriptsize}
\rotate
\tablecaption{Non-members \label{table:hecto_targets_nonmem}}
\tablewidth{0pt}
\tablehead{
\colhead{2MASS\_id} & \colhead{ID} & \colhead{J} & \colhead{$(J-H)$} & \colhead{$(H-K)$} 
& \colhead{$[3.6-4.5]$} & \colhead{$[5.8-8.0]$} & \colhead{$V_{rad}$} & \colhead{$\Delta V_{rad}$} & \colhead{R} & \colhead{S} & \colhead{$EW_{c}$} & \colhead{$FW_{10\%}$} & \colhead{$H\alpha$} & \colhead{CCF} &\colhead{NOB} \\
}
\startdata
0533283-045549	& F11-ap48	& 13.27	& 0.89	& 0.24	& 0.04	& 0.03	& -22.07	& 0.94	& 8.9	& 0.46	& 0.0	& 0	& AE	& c	& 1\\
0533289-050930	& F22-ap46	& 12.16	& 1.04	& 0.30	& 0.05	& 0.37	& 106.39	& 0.53	& 13.2	& 0.61	& 0.0	& 0	& R	& c	& 1\\
0533298-052735	& F21-ap66	& 13.31	& 0.88	& 0.25	& 0.02	& 0.53	& -20.75	& 0.47	& 19.2	& 0.74	& 0.0	& 0	& AE	& c	& 1\\
0533333-043918	& F11-ap67	& 12.89	& 0.91	& 0.27	& 0.00	& 0.00	& -6.00		& 0.62	& 11.5	& 0.49	& 0.0	& 0	& AE	& c	& 1\\
0533363-050140	& F11-ap35	& 13.29	& 0.80	& 0.21	& 0.00	& 0.00	& --		& --	& 2.1	& 0.17	& 0.0	& 0	& AE	& w,l	& 1\\
0533390-055434	& F31-ap72	& 12.59	& 1.62	& 0.40	& -0.02	& 0.04	& 73.01		& 0.38	& 19.7	& 0.70	& 0.0	& 0	& AE	& c	& 1\\
0533415-050934	& F22-ap50	& 13.19	& 0.96	& 0.24	& 0.08	& -0.03	& -11.32	& 0.81	& 9.8	& 0.58	& 0.0	& 0	& AE	& c	& 1\\
0533445-045930	& F11-ap31	& 12.62	& 1.05	& 0.23	& -0.07	& 0.11	& 12.70		& 0.34	& 20.5	& 0.72	& 0.0	& 0	& AE	& c	& 1\\
0533497-043329	& F11-ap73	& 13.47	& 0.98	& 0.25	& 0.00	& 0.00	& 46.59		& 0.89	& 7.4	& 0.34	& 0.0	& 0	& AE	& c	& 1\\
0533514-043655	& F11-ap78	& 13.37	& 1.25	& 0.34	& 0.00	& 0.00	& -6.63		& 10.17	& 2.9	& 0.22	& 0.0	& 0	& AE	& d,l	& 1\\
0533520-050655	& F21-ap93	& 12.76	& 0.92	& 0.27	& 0.02	& -0.06	& -9.05		& 0.38	& 12.7	& 0.56	& 0.0	& 0	& N	& c	& 2\\
0533524-044400	& F11-ap64	& 12.12	& 1.20	& 0.30	& -0.01	& 0.17	& -5.25		& 0.34	& 21.6	& 0.72	& 0.0	& 0	& AE	& c	& 1\\
0533570-054210	& F22-ap235	& 12.08	& 1.66	& 0.44	& -0.08	& 0.16	& 112.78	& 0.46	& 15.2	& 0.68	& 0.0	& 0	& R	& c	& 1\\
0533579-060524	& F31-ap53	& 13.10	& 0.80	& 0.24	& 0.06	& 0.04	& 38.64		& 0.76	& 10.1	& 0.60	& 0.0	& 0	& AE	& c	& 1\\
0534018-055941	& F31-ap69	& 13.17	& 1.24	& 0.32	& 0.00	& 0.12	& 70.32		& 0.37	& 19.2	& 0.71	& 0.0	& 0	& AE	& c	& 1\\
0534040-061329	& F31-ap44	& 12.70	& 0.80	& 0.20	& 0.01	& 0.17	& 4.17		& 0.57	& 15.1	& 0.75	& 0.0	& 0	& AE	& c	& 1\\
0534057-062851	& F31-ap22	& 11.77	& 1.31	& 0.38	& 0.00	& 0.00	& 113.63	& 0.41	& 18.0	& 0.68	& 0.0	& 0	& R	& c	& 1\\
0534071-054925	& F31-ap82	& 13.17	& 1.12	& 0.35	& 0.02	& -0.11	& 38.22		& 0.75	& 10.3	& 0.49	& 0.0	& 0	& AE	& c	& 1\\
0534073-043231	& F11-ap82	& 12.86	& 0.91	& 0.29	& 0.00	& 0.00	& -13.51	& 0.74	& 10.3	& 0.49	& 0.0	& 0	& AE	& c	& 1\\
0534080-060924	& F31-ap41	& 13.46	& 1.03	& 0.29	& 0.01	& 0.10	& 1.20		& 0.45	& 18.9	& 0.74	& 0.0	& 0	& AE	& c	& 1\\
0534095-050111	& F11-ap21	& 13.18	& 1.01	& 0.25	& 0.04	& -0.14	& -4.65		& 0.60	& 10.3	& 0.57	& 0.0	& 0	& AE	& c,dv	& 2\\
0534098-042819	& F11-ap87	& 12.08	& 0.69	& 0.21	& 0.00	& 0.00	& 53.92		& 0.60	& 13.4	& 0.67	& 0.0	& 0	& AE	& c	& 1\\
0534104-051020	& F22-ap54	& 12.97	& 0.95	& 0.31	& 0.06	& 0.43	& -40.46	& 0.91	& 8.4	& 0.53	& 0.0	& 0	& R	& c	& 1\\
\enddata
\tablecomments{Hectochelle targets in ONC found to be non-members based on measured RV value
or lack of detected H$\alpha$ emission. Stars are listed regardless of R value, but no velocity
is displayed in case of very low R or undefined CCF.
--- For the full table see the electronic version of teh Jurnal or contact
the authors.\\
~~~
\textbf{2MASS id} --- 2MASS identification number 
(truncated RA and DEC coordiantes as: HHMMSSS+DDMMSS);~~
\textbf{ID} --- internal identification number, specifying the field
(see Table \ref{table:obs}) and aperture of observation.  The letter \textit{F} 
means the first run, and therefore 2MASS based selection, the letter \textit{S}
notes IRAC based selction and observation taken during the 2nd run;~~
\textbf{J} --- 2MASS J magnitude;~~
\textbf{(J-H)} --- 2MASS $(J-H)$ color index;~~ 
\textbf{(H-K)} --- 2MASS $(H-K)$ color index;~~
\textbf{$3.6-4.5$} ---  IRAC short wavelength color index;~~
\textbf{$5.8-8.0$} ---  IRAC long wavelength color index;~~
\textbf{$V_{rad}$} --- measured heliocentric radial velocity, in \kms;~~
\textbf{$V_{rad}$} --- \textit{xcsao} error estimate for $V_{rad}$, in \kms;~~
\textbf{R} --- R value of cross correlation (see text for details);~~
\textbf{S} --- height of the CCF peak;~~
\textbf{$EW_c$} --- absolute value of equivalent width, measured on the corrected
\halp~ profile (see text for details);~~
\textbf{$FW_{10\%}$} --- full width of \halp~ profile at 10\% level of the corrected
maximum (see text for details);\\
~~~
\textbf{H$\alpha$} ---  notes on the \halp~ emission profile:
 \textbf{D} --- resolved or unresolved double gaussian profile, 
  no excess H-alpha emission (likley non-TTS);~~
 \textbf{R} --- obviously shifted H-alpha absorption : high RV stars;~~
 \textbf{X} --- very wide H-alpha absorption with nebular emissions;~~
 \textbf{AE} --- stellar absorption and nebular emission together;~~
 \textbf{NS} --- very strong nebular component component, no wings, no asymmetry;~~
~~~
\textbf{CCF} --- notes on the cross-correlation function:
 \textbf{w} -- wide peak, results large errors in RV;
 \textbf{l} -- very low peak, but still sticks out from surrounding;
 \textbf{u} -- almost undefined, very low/wide peak;
 \textbf{n} -- very noisy (several local peaks, almost as high as the one picked);
 \textbf{s} -- side lobed, could be a spectroscopic binary;
 \textbf{d} -- double peak, spectroscopic binary;
 \textbf{dv} -- RV measured more than once, found different velocities, could be binary;
 \textbf{r} -- template spectra with highest peak in cross-correlation resulted non-realistic velocity,
so the template resulting in highest R value was used instead to determine velocity;
 \textbf{c} -- clear, well-isolated peak;
 \textbf{?} -- uncertainty in assigning the given note;\\
~~~
\textbf{NOB} --- number of observations}
\end{deluxetable}

\end{document}